\documentstyle[prd,aps,preprint,tighten,eqsecnum,epsf]{revtex}
\newcommand{\parasmall}{_{\stackrel{\parallel}{\,}}}
\begin{document}
\preprint{{\tt gr-qc/9612020}} 

\title{Differential Forms and Wave Equations for General 
Relativity\footnote{Preprint TUW-96-09 (revised). 
Research supported by the ``Fonds zur F\"{o}r\-der\-ung 
der wis\-sen\-schaft\-lich\-en For\-schung'' 
in Austria (Lise Meitner Fellowship M-00182-PHY and FWF Project
10.221-PHY).}} 

\author{Stephen R. Lau\footnote{email address: 
{\tt lau@tph16.tuwien.ac.at}}}

\address{Institut f\"{u}r Theoretische Physik,\\ Technische 
Universit\"{a}t Wien,\\ Wiedner Hauptstra\ss e 8-10,\\ A-1040 
Wien, \"{O}sterreich}

\maketitle
\begin{abstract}
Recently, Choquet-Bruhat and York and Abrahams, Anderson,
Choquet-Bruhat, and York ({\sc aacy}) have cast the 3+1 evolution 
equations of general relativity 
in gauge-covariant and causal ``first-order symmetric 
hyperbolic form,'' thereby cleanly separating
physical from gauge degrees of freedom in the Cauchy problem
for general relativity. A key ingredient in their construction 
is a certain wave equation which governs the light-speed 
propagation of the extrinsic curvature tensor. 
Along a similar 
line, we construct a related wave equation which, as the key 
equation in a system, describes vacuum general relativity. 
Whereas the approach of {\sc aacy} is based on tensor-index 
methods, the present formulation is written solely in the 
language of differential forms. Our approach starts with 
Sparling's tetrad-dependent differential forms, and our wave 
equation governs the propagation of Sparling's 2-form, which 
in the ``time-gauge" is built linearly from the ``extrinsic 
curvature 1-form.'' The tensor-index version of our wave 
equation describes the propagation of (what is essentially) the 
Arnowitt-Deser-Misner gravitational momentum.  
\end{abstract}
\vfill
Vienna, December 1996
\newpage
\section*{Introduction}

The 3+1 formulation of general relativity views spacetime as 
the time history of the geometry associated with a spacelike 
hypersurface $\Sigma$. In the differential-forms ({\sc df}) 
version\footnote{Quite a number of reviews of 
``form relativity'' exist. An excellent choice is the 
discussion found in Ref.~\cite{canonical} by Isenberg and 
Nester.} of 3+1 relativity, the (co)triad 1-form 
$e^{{\rm a}}$ and the ``extrinsic curvature 1-form''  
$K^{{\rm a}}$ are taken as the basic geometrodynamical 
variables of interest.\cite{canonical} In this description 
initial data sets $(e^{{\rm a}} , K^{{\rm a}})$ are constrained 
by seven point-wise relations, {\sc df} versions of 
the familiar Hamiltonian, momentum, and rotation constraints; 
and the Arnowitt-Deser-Misner 
({\sc adm})\cite{ADM,canonical} equations of motion for $e^{{\rm a}}$ 
and $K^{{\rm a}}$ govern the dynamic evolution of the system. 
These evolution equations preserve the constraints, and, 
moreover, maintain complete spatial covariance by means of 
an arbitrary shift vector $\beta^{j}$ and an arbitrary 
skew-symmetric rotation matrix $\phi_{{\rm a}{\rm b}}$ 
(covariance under spatial coordinate transformations is not 
an issue in the {\sc df} language). However, as is 
well-known these equations do not manifest mathematically the 
propagation of physical degrees of freedom along the 
lightcone; they do not give rise to a physical wave equation. 
Therefore, in 3+1 form relativity (as in its pure-index metric 
counterpart) gauge and physical variables remain intertwined.

Recently, Choquet-Bruhat and York \cite{YCBandJWY} and 
Abrahams, Anderson, Choquet-Bruhat and York \cite{AACY} 
(hereafter, we refer to both citations jointly as {\sc aacy}) 
have cast the 3+1 evolution equations for metric general 
relativity in gauge-covariant and causal 
``first-order symmetric hyperbolic form'' ({\sc fosh} form), 
thereby cleanly separating physical from gauge degrees of 
freedom in the Cauchy problem for general relativity. A key
ingredient in their construction is a certain wave equation 
which governs the light-speed propagation of the extrinsic 
curvature tensor $K_{ij}$. Along a similar line, we construct 
a related wave equation which, as the key equation in a 
system, describes general relativity. (Here we only consider 
vacuum, although we expect the inclusion of matter to be 
straightforward enough.) Whereas the 
approach of {\sc aacy} is based on tensor methods, 
the present formulation is written solely in the language of 
differential forms. Our approach starts with Sparling's 
tetrad-dependent differential forms,\cite{Madore,Goldberg} 
and our wave equation 
governs the propagation of 
Sparling's 2-form, which in the ``time-gauge" is built 
linearly from the extrinsic curvature 1-form. The 
tensor-index version of our wave equation describes the 
propagation of (what is essentially) the 
Arnowitt-Deser-Misner ({\sc adm}) gravitational momentum. 
As with {\sc aacy}'s ``Box K'' equation, to be
proper our {\sc df} wave equation requires the implementation of 
a harmonic time slicing (or suitable generalization thereof) in
order to handle the time re-parameterization invariance of 
Einstein's theory. 
In this paper we only derive the described wave equation.  
We do not discuss the possibility of constructing {\sc fosh} systems 
for Einstein's theory related to the one given by {\sc aacy}. 
(We hope to return to this issue later.) Our goal is to understand 
the geometry of the {\sc aacy} formalism in the {\sc df} 
language. Although there is no doubt in
our mind that the original pure-index formulation of {\sc aacy} is the
superior wave-equation version of Einstein's theory, one might 
expect a {\sc df} construction to show distinct advantages in some 
theoretical contexts. 

{\sc aacy} obtain their wave equation for $K_{ij}$ by (i) taking a 
further ``time derivative'' of the equation of motion for 
$K_{ij}$, and then (ii) subtracting from 
the resulting expression certain 
spatial covariant derivatives of the momentum constraint 
(which has $\beta^{j}$ as its corresponding Lagrange multiplier). 
Now, the standard {\sc df} version of 3+1 relativity 
is, in essence, a triad theory, requiring the existence of three 
auxiliary phase-space constraints (packaged into one name, the 
``rotation constraint,'' with corresponding Lagrange 
multiplier $\phi_{{\rm a}{\rm c}}$). 
In our wave-equation construction, 
we proceed in a similar fashion to {\sc aacy}. We also take the time 
derivative of an equation of motion and then subtract from
the resulting
expression certain spatial covariant derivatives of the momentum 
constraint. However, we find the need to also subtract from this
time derivative certain spatial covariant derivatives of 
the rotation constraint. To us this seems to be natural 
generalization of the {\sc aacy} method, necessary to handle the extra
gauge freedom present in the triad theory.
It is well-known that geometrically the rotation
constraint is related to a certain piece of the spacetime 
torsion; and, therefore, in this paper we keep track of this 
very piece. Of course, this has consequences for the usual 
symmetries enjoyed by the Riemann, Ricci, and Einstein tensors 
(and {\sc df} versions thereof). 
However, we wish to stress here that we do not consider 
Einstein-Cartan theory in this paper. 
``At the end of the day,'' we set all torsion 
equal to zero. We keep track of torsion in the beginning only as a 
rigorous way to handle the issue of the rotation constraint, which 
is {\em always} present 
in a triad version of Einstein's theory.\cite{canonical,Goldberg,Ashtekar}

The organization of this paper is as follows. In a preliminary 
section, $\S$ I, we fix our conventions and collect the geometric 
tools necessary for our central discussion. In particular, we 
quickly review the 3+1 decomposition of spacetime differential 
forms and introduce a certain evolution operator $\hat{D}_{0}$, 
which may be regarded as a (triad) generalization of the evolution 
operator $\hat{\partial}_{0}$ central to the work of {\sc aacy}. The 
reader may wish to simply peruse the preliminary section, and if
necessary return to it later for more detailed study.  
$\S$ II deals with geometry of vacuum spacetimes as described by
differential forms, and it lays the groundwork for our wave-equation
formulation of general relativity. $\S$ III introduces Sparling's
forms, using them to enact a {\sc df} version of the standard 3+1
decomposition of spacetime geometry, while $\S$ IV combines the
results of $\S$ II and III to write down a system of {\sc df} equations
for general relativity, one equation in the system being the 
promised wave equation. We demonstrate the equivalence between
our {\sc df} system and the original Einstein equations in $\S$ V. 
The concluding section provides some ``instructions'' for 
rewriting our results in tensor-index form. Finally, we have 
included as appendices a handful of technical derivations 
necessary for our central discussion. 

\section{Preliminaries}

\subsection{Notation and Conventions}
 
Consider a spacetime $\cal M$, equipped with Lorentz-signature 
metric $g_{\mu\nu}$, which is foliated by a collection of 
spacelike 3-dimensional slices $\Sigma$. We use $\Sigma$ both 
to denote the foliation and to represent a generic slice.
Since we are concerned only with local field equations in 
this work, we do not further specify the global properties of 
$\Sigma$. The timelike, future-pointing, unit, hypersurface 
normal of the $\Sigma$ foliation is $u^{\mu}$. Respectively, 
$h_{ij}$ and $K_{ij}$ denote the intrinsic metric and extrinsic 
curvature tensor associated with the $\Sigma$ foliation.

We shall use two types of spacetime frames in this work. 
One is the foliation-adapted ``quasi-coordinate'' frame 
$e_{\mu}$ (with coframe $e^{\mu}$) used by {\sc aacy}
\begin{eqnarray}
e^{0} = {\rm d}t 
& \hspace{1cm} & 
e_{0} = \partial/\partial t 
- \beta^{j} \partial/\partial x^{j} 
\nonumber \\
& & \label{qcframe}\\
e^{j} = {\rm d}x^{j} + \beta^{j} {\rm d}t 
& &
e_{j} = \partial/\partial x^{j} 
{\,} . 
\nonumber
\end{eqnarray}
Greek letters $\{\mu, \nu, \lambda, \cdots \}$ run over 
$0,1,2,3$, but $0$ is not $t$. Here $N$ and $\beta^{k}$ 
are respectively the standard {\em lapse function} and 
{\em shift vector} associated with the $\Sigma$ foliation. 
Lowercase Latin letters $\{i,j,k, \cdots\}$, taking the 
values $1,2,3$, represent the space legs of the 
quasi-coordinate frame and are also $\Sigma$ coordinate 
indices. We use $\partial_{0}$ to represent $e_{0}$ as 
a Pfaff derivative,\cite{AMP} so for some scalar function 
$f$ we write $\partial_{0}f$ to mean 
$\dot{f} - \beta^{j} \partial f/\partial x^{j}$.

The second type of frame we consider is the foliation-adapted 
(or {\em time-gauge}) tetrad $e_{{A}}$ (with coframe
$e^{{A}}$):
\begin{eqnarray}
e^{\bot} = N e^{0}  
& \hspace{1cm} & 
e_{\bot} = N^{-1} e_{0} \nonumber \\
& & \label{tgtetrad}\\
e^{{\rm a}} = e^{\rm a}\,_{j} e^{j} 
& &
e_{{\rm a}} = e_{{\rm a}}\,^{j} e_{j} 
{\,} . 
\nonumber
\end{eqnarray}
Notice that $e_{\bot}{}^{\mu}$ is the $\Sigma$ normal $u^{\mu}$. 
Capital Latin letters $\{{A},{B}, {C},\cdots\}$ represent 
tetrad indices and run over $\bot, \hat{1},\hat{2}, \hat{3}$, 
while block lowercase letters, typically from the first part of 
the alphabet, $\{{\rm a}, {\rm b}, {\rm c},\cdots\}$ represent 
$\Sigma$ triad indices and run over $\hat{1},\hat{2},\hat{3}$. 
The set of $e_{{\rm a}}\,^{j}$ comprise an orthonormal triad 
of vector fields on $\Sigma$. Orthonormality implies that 
$h_{ij} e_{{\rm a}}\,^{i} e_{{\rm b}}\,^{j}
= \delta_{{\rm a}{\rm b}}$, with $\delta_{{\rm a}{\rm b}} = 
\delta^{{\rm a}{\rm b}} = diag(1,1,1)$. We can represent the
$\Sigma$ metric as $h_{ij} = \delta_{{\rm a}{\rm b}} 
e^{{\rm a}}\,_{i} e^{{\rm b}}\,_{j}$.
 
\subsection{3+1 Splitting of Forms and the First
Cartan Structure Equation}

We may split a general spacetime differential form $\Psi$ 
into pieces normal and tangential to the $\Sigma$ foliation,
\begin{equation}
\Psi = {}^{\parasmall}\! \Psi + 
e^{\bot} \wedge \Psi_{\bot}
{\,} ,
\label{receipe}
\end{equation}
where the $i_{\bot} \equiv i_{u}$ is the ``inner derivative'' 
with $u$, also known as the vector-form  ``hook" with 
$u$.\citation{GandS} The hook of a spatial form with the 
timelike $\Sigma$ normal $u = e_{\bot}$ vanishes. Both 
${}^{\parasmall}\!\Psi$ and $\Psi_{\bot}$ are spatial forms.
Let us take a special look at {\em spacetime} forms which
happen to be spatial to start with. For the sake of
definiteness only, consider a spatial 2-form $\psi 
= {}^{\parasmall}\! \psi$. In the index notation and with 
the quasi-coordinate basis, the only non-vanishing components 
of $\psi_{\mu\nu}$ are $\psi_{ij}$, so that 
$\psi = \frac{1}{2} \psi_{ij} e^{i} \wedge e^{j}$. Note that, 
while $\psi$ is purely spatial, it is still a {\em spacetime} 
form; we have not pulled it back to a {\em particular} $\Sigma$ 
slice. This is an advantage gained by using the 
quasi-coordinate frame: all forms (in fact, all tensors) 
which are parallel to the $\Sigma$ foliation 
{\em in spacetime} may be represented with $\Sigma$ 
indices in the index notation. As another example, 
take $e^{{\rm a}} = {}^{\parasmall}\! e^{{\rm a}}$, which 
has non-vanishing components $e^{{\rm a}}\,_{j}$.
 
Now suppose $\Psi$ is a 3-form or less. We have the following 
decomposition of ${\rm d}\Psi$ with respect to the $\Sigma$ 
foliation:\cite{canonical}
\begin{equation}
  {\rm d}\Psi 
= d{}^{\parasmall}\! \Psi 
+ N^{-1} e^{\bot} \wedge 
  \left[ \hat{\partial}_{0} 
  {}^{\parasmall}\!\Psi 
- d(N \Psi_{\bot}) \right]
{\,} , 
\label{decomp}
\end{equation}
where $\rm d$ is the $\cal M$ exterior derivative and $d$ is the 
$\Sigma$ exterior derivative. Also, our ``time-evolution operator'' 
is $\hat{\partial}_{0} \equiv \pounds_{e_{0}}$, with ${\pounds}$ 
representing $\cal M$ Lie differentiation.\footnote{It is precisely 
by choosing to take the Lie derivative along $e_{0}{}^{\mu}$, that 
we may use {\em spacetime} $\pounds$ in (\ref{decomp}) in place of 
the ``intrinsic $\Sigma$ Lie derivative generalized to take care 
of derivatives in directions off of the $\Sigma$ hypersurfaces'' 
discussed in Ref. \cite{canonical}. Had we chosen to take the Lie 
derivative along $u^{\mu}$, as is often done, then we would
have needed to introduce such a generalized $\Sigma$ Lie 
derivative.} Note that $\hat{\partial}_{0}$ is not the Pfaff 
derivative $\partial_{0} 
= \partial/\partial t - \beta^{k} \partial/\partial x^{k}$, 
although it agrees with $\partial_{0}$ when acting on scalars.
In the index notation, the action of $\hat{\partial}_{0}$ on a 
$\Sigma$ form, for example 
$\psi_{\lambda} = \psi_{i} e^{i}{}_{\lambda}$, is given by 
$\pounds_{e_{0}}\psi_{\lambda} = e^{i}{}_{\lambda}(\dot{\psi}_{i} 
- L_{\beta}\psi_{i})$, with $L$ denoting intrinsic $\Sigma$ Lie 
differentiation. Thus, 
while working (in the index notation) with 
$\Sigma$ tensors carrying {\em spatial} indices $(i,j,k,\cdots)$, 
it is consistent to set $\hat{\partial}_{0} 
= \partial/\partial t - L_{\beta}$.

Applying the exterior derivative decomposition (\ref{decomp})  
twice to the formula 
${\rm d}^{2} \psi = 0$, where now $\psi = {}^{\parasmall}\! \psi$ 
is an arbitrary spatial form, one proves that the following 
commutator vanishes:
\begin{equation}
[\hat{\partial}_{0} , d]\psi \equiv \hat{\partial}_{0} d \psi
- d \hat{\partial}_{0} \psi = 0 {\,} .
\label{formcommutator}
\end{equation}  
In index notation (\ref{formcommutator}) can be translated 
into the statement that $\hat{\partial}_{0}$ commutes with 
$\partial_{k}$ partial differentiation. However, for the 
Pfaff derivative $[\partial_{0}, \partial_{k}] \neq 0$. 

Consider the first structure equation of Cartan,
\begin{equation}
T^{{A}} 
= {\cal D} e^{{A}} \equiv {\rm d}e^{{A}} 
+ \Gamma^{{A}}\,_{{B}} \wedge e^{{B}} 
{\,}  , \label{Cartanfse}
\end{equation}
where 
$T^{{A}} = \frac{1}{2} e^{A}{}_{\lambda} 
T^{\lambda}{}_{\mu\nu} e^{\mu} \wedge e^{\nu}$ 
is the torsion 2-form, the $\Gamma^{A}{}_{B}$ are
the 1-forms specifying the spacetime connection
with respect to the time-gauge tetrad (\ref{tgtetrad}),
and ${\cal D}$ is the spacetime exterior covariant 
derivative. 
We shall assume that $T^{\bot}{}_{ij}$ 
is the only piece of the spacetime torsion tensor which is not 
manifestly zero; hence we only need to consider the $T \equiv 
{}^{\parasmall}\! T^{\bot}$ piece of the torsion 2-form.
 
The ${}^{\parasmall}$ projections of the 
$(\bot,{\rm a})$ values of (\ref{Cartanfse}) tell us that
\begin{eqnarray}
0 & = & 
D e^{{\rm a}}  
\equiv d e^{{\rm a}} 
+ \omega^{{\rm a}}\,_{{\rm b}} 
\wedge e^{{\rm b}} 
\label{cartan1} 
\eqnum{\ref{cartan1}a} \\
& &  \nonumber \\
T & = & K_{{\rm a}{\rm b}} e^{{\rm a}} \wedge e^{{\rm b}} 
{\,} , \eqnum{\ref{cartan1}b} 
\addtocounter{equation}{1}
\end{eqnarray}
where $\omega^{{\rm a}}{}_{{\rm b}} =
{}^{\parasmall}\!\Gamma^{{\rm a}}{}_{{\rm b}}$ are the $\Sigma$
triad connection 1-forms and $D$ is the $\Sigma$ exterior
covariant derivative.
The top equation says that the 
intrinsic connection on each slice $\Sigma$ is torsion-free.
The second equation follows because (i)
$e^{\bot} \wedge {\rm d}e^{\bot} = 0$ and (ii) 
the triad components of the $\Sigma$ extrinsic curvature 
tensor are defined in the time-gauge by 
$K_{{\rm a}{\rm b}} \equiv - \Gamma_{{\rm a}\bot{\rm b}}$.
Therefore, setting $T$ equal to zero enforces the symmetry 
of $K_{ij}$. Taking the $i_{\bot}$ inner derivative of 
(\ref{Cartanfse}), one finds that 
\begin{eqnarray}
0 & = & \hat{\partial}_{0} e^{{\rm a}} + 
\Gamma^{{\rm a}}\,_{{\rm b}0}
e^{{\rm b}}
+ N K^{{\rm a}}\,_{{\rm b}} e^{{\rm b}} 
\label{cartan2} \eqnum{\ref{cartan2}a} \\
& & \nonumber \\
0 & = & - a_{{\rm b}} + \Gamma^{\bot}\,_{{\rm b}\bot}  
\eqnum{\ref{cartan2}b}
{\,} ,
\addtocounter{equation}{1}
\end{eqnarray}
with $a_{{\rm b}} \equiv e_{{\rm b}}[\log N]$. It
follows that the $a^{{\rm b}}$ are the triad components 
of the acceleration of the $\Sigma$ foliation normal 
$u^{\mu}$.\cite{canonical}
Upon inspection of (\ref{cartan2}a), we define 
a generalized version $\hat{D}_{0}$ of the evolution 
operator $\hat{\partial}_{0}$ by
\begin{equation}
  \hat{D}_{0} e^{{\rm a}} 
  \equiv \hat{\partial}_{0} e^{{\rm a}} 
  + \Gamma^{{\rm a}}\,_{{\rm b}0} 
  e^{{\rm b}}
{\,} ,
\label{definitionD_0}
\end{equation}
and will shortly extend its action to other forms and 
tensors.

\subsection{Triad-covariant Evolution Operator}

The advantage gained by using the simple evolution operator
$\hat{\partial}_{0}$ is well-known.\cite{sources} 
Differentiation by $\hat{\partial}_{0}$ preserves the 
spatial character of any 
spatial form or spatial tensor. Moreover, $\hat{\partial}_{0}$ 
defines a good ``time axis'' as it is
orthogonal to the {\em spacelike} $\Sigma$ foliation.
Our generalization of $\hat{\partial}_{0}$, the degree-zero 
evolution operator $\hat{D}_{0}$ introduced
in (\ref{definitionD_0}), is covariant with respect 
to triad or ``internal" rotations. Now, in fact, the 
canonical formulation of triad gravity identifies
the anti-commuting Lagrange parameter 
$\phi^{{\rm a}}\,_{{\rm b}}$ associated with the rotation 
constraint as $- \Gamma^{{\rm a}}\,_{{\rm b}0} = 
- N \Gamma^{{\rm a}}\,_{{\rm b}\bot}$. 
The connection coefficients $\Gamma^{{\rm b}}\,_{{\rm a}0}$
describe the rotation of the triad as it is parallel 
transported along the integral 
curves of $e_{0}{}^{\mu} = Nu{}^{\mu}$. We generalize
the action of $\hat{D}_{0}$ to other triad tensor-valued
forms in the obvious way. For example, consider a 
triad tensor-valued 0-form $f_{{\rm a}}{}^{{\rm b}}$.
On such an object, the direct way of expressing the action of 
$\hat{D}_{0}$ would be
\begin{equation}
\hat{D}_{0} f_{{\rm a}}{}^{{\rm b}} =
D_{0} f_{{\rm a}}{}^{{\rm b}} \equiv
\partial_{0}f_{{\rm a}}{}^{{\rm b}} - f_{{\rm c}}{}^{{\rm b}} 
\Gamma^{{\rm c}}{}_{{\rm a}0} + f_{{\rm a}}{}^{{\rm c}} 
\Gamma^{{\rm b}}{}_{{\rm c}0}
{\,} . \label{firstway}
\end{equation} 
Notice that on 0-forms $\hat{D}_{0}$ reduces to $D_{0}$.
If we had taken $f_{{\rm a}}{}^{{\rm b}}$ as a tensor-valued
1-form $f_{{\rm a}}{}^{{\rm b}}{}_{k} e^{k}$, then we would
have had to use the operator $\hat{\partial}_{0}$ in the
first term on the right-most side of (\ref{firstway}). In other
words, the ``hat'' does not see triad indices, rather it 
sees whether or not the differentiated object is a form of
degree higher than zero. Equation (\ref{cartan2}a) ensures 
that we have
\begin{equation}
\hat{D}_{0} e^{{\rm a}} = -
N K^{{\rm a}} {\,} , \label{triadevolution}
\end{equation}
with the extrinsic curvature appearing as a one-form 
$K^{{\rm a}} = K^{{\rm a}}{}_{j}e^{j}$, as the evolution 
rule for the (co)triad. 

Let us address some technical points concerning how to
use the operator $\hat{D}_{0}$ in index notation. 
First, when working in index notation, assume the following 
evolution rules for the cotriad and triad:
\begin{eqnarray}
\hat{D}_{0} e^{{\rm a}}{}_{j} 
& \equiv & \hat{\partial}_{0} e^{{\rm a}}{}_{j}
+ e^{{\rm b}}{}_{j} \Gamma^{{\rm a}}{}_{{\rm b}0} 
= - N K^{{\rm a}}{}_{j}
\label{indextriadevolution} 
\eqnum{\ref{indextriadevolution}a} \\
\hat{D}_{0} e_{{\rm a}}{}^{j} 
& \equiv & \hat{\partial}_{0} e_{{\rm a}}{}^{j} 
- e_{{\rm b}}{}^{j} \Gamma^{{\rm b}}{}_{{\rm a}0} 
= N e_{{\rm b}}{}^{j} K^{{\rm b}}{}_{{\rm a}}
{\,} . 
\eqnum{\ref{indextriadevolution}b}
\addtocounter{equation}{1}
\end{eqnarray}
Second, assume that $\hat{D}_{0}$ obeys the Leibnitz rule
and reduces to $\hat{\partial}_{0}$ when acting on tensors
without triad indices. As an example of how to proceed in 
index notation, consider again our triad tensor-valued 
0-form, but assume that its triad components 
$f_{{\rm a}}{}^{{\rm b}} = f_{i}{}^{j} e_{{\rm a}}{}^{i} 
e^{{\rm b}}{}_{j}$ are obtained by ``soldering'' all of the free 
indices of a $\Sigma$ tensor $f_{i}{}^{j}$. 
One can also obtain the action of 
$\hat{D}_{0}$ on $f_{{\rm a}}{}^{{\rm b}}$ by letting it act 
on $f_{i}{}^{j} e_{{\rm a}}{}^{i} e^{{\rm b}}{}_{j}$ and then 
using the Leibnitz rule. The result is 
\begin{equation}
\hat{D}_{0} f_{{\rm a}}{}^{{\rm b}} =
e_{{\rm a}}{}^{i} e^{{\rm b}}{}_{j}
\hat{\partial}_{0} f_{i}{}^{j} 
+ N f_{{\rm c}}{}^{{\rm b}} K^{{\rm c}}{}_{{\rm a}} 
- N f_{{\rm a}}{}^{{\rm c}} K^{{\rm b}}{}_{{\rm c}}
{\,} .
\label{secondway}
\end{equation}
Of course, (\ref{firstway}) and (\ref{secondway}) agree. 
By first writing $\delta_{{\rm a}{\rm b}} 
= h_{ij} e_{{\rm a}}{}^{i} e_{{\rm b}}{}^{j}$ 
and $\epsilon_{{\rm a}{\rm b}{\rm c}} 
= \epsilon_{ijk} e_{{\rm a}}{}^{i} 
e_{{\rm b}}{}^{j} e_{{\rm c}}{}^{k}$, where $\epsilon_{ijk}$
is the $\Sigma$ permutation tensor fixed by 
$\epsilon_{123} = \sqrt{h}$, one can use the result
$\hat{\partial}_{0} h_{ij} = - 2N K_{(ij)}$ to show that
$\hat{D}_{0}$ annihilates both $\delta_{{\rm a}{\rm b}}$ and
$\epsilon_{{\rm a}{\rm b}{\rm c}}$. That $\hat{D}_{0}$ kills
these particular triad tensors can also be shown with formulae
like (\ref{firstway}) along with the fact that the 
$\Gamma_{{\rm a}{\rm b}0}$ are antisymmetric in their triad 
indices.

We have already seen that $\hat{\partial}_{0}$ 
and $d$ commute when acting on purely spatial forms. What
about $\hat{D}_{0}$ and $D$? ($D$ is the $\Sigma$
exterior covariant derivative.) Appendix B proves 
that for any triad-valued 
$\Sigma$ form $\psi^{{\rm a}} = 
{}^{\parasmall}\!\psi^{{\rm a}}$, we have the following 
commutation rule:
\begin{equation}
[\hat{D}_{0}, D]\psi^{\rm a} \equiv (\hat{D}_{0} D - 
D\hat{D}_{0}) \psi^{{\rm a}}
=  N\left( \Re^{{\rm a}}\,_{{\rm b}\bot} 
+ a^{{\rm a}} K_{{\rm b}} 
- a_{{\rm b}} K^{{\rm a}}\right) 
\wedge \psi^{{\rm b}}\, ,
\label{commutatorhatD_0andD}
\end{equation}
where  $\Re^{{\rm a}}{}_{{\rm b}\bot} = i_{\bot}
\Re^{{\rm a}}{}_{{\rm b}}$ and the 
$\Re^{{\rm a}}{}_{{\rm b}}$ comprise the triad block of the 
matrix-valued curvature two-form of Cartan, 
$\Re^{{A}}{}_{{B}}
= \frac{1}{2}\Re^{{A}}\,_{{B}\mu\nu} e^{\mu} \wedge
e^{\nu}$. Here the $\Re^{{A}}\,_{{B}\mu\nu} = e^{{A}}{}_{\alpha}
e_{{B}}{}^{\beta} \Re^{\alpha}{}_{\beta\mu\nu}$ 
are the mixed components of the spacetime Riemann tensor.


\section{Einstein gravity in terms of differential forms}

Working within the Cartan framework of tensor-valued differential 
forms, in this section we collect a handful of standard 
results concerning the geometry of vacuum Einstein spacetimes.
We carefully examine the role played by the residual torsion 
in our formalism, in particular deriving several identities
which will be used later on to demonstrate that the residual 
torsion may safely removed. 

\subsection{Einstein 3-form}

Consider the Einstein tensor 
$G_{\beta{A}} \equiv e_{{A}}{}^{\alpha} G_{\beta\alpha}$ 
written with respect to a mixed basis. Here and it what 
follows we pay strict attention to index ordering.
Since we are keeping track of certain components of the 
spacetime torsion, the Einstein {\em tensor} is not 
symmetric {\em a priori} in our formalism. We will 
work with the {\em Einstein 3-form}
\begin{equation} 
   {\sf G}_{{A}} \equiv
   {\textstyle \frac{1}{6}}
   G_{\kappa{A}} 
   \epsilon^{\kappa}{}_{\alpha\beta\lambda} 
   e^{\alpha} \wedge e^{\beta} 
   \wedge e^{\lambda} 
{\,} .
\label{Einstein3form}
\end{equation}
Clearly, ${}^{*} {\sf G}_{{A}} = G_{\beta{A}}e^{\beta}$, 
where $*$ denotes the Hodge-duality defined by the 
${\cal M}$ metric $g_{\mu\nu}$. 
In (\ref{Einstein3form}) the ${\cal M}$
permutation tensor $\epsilon_{\lambda\mu\nu\sigma}$ is
chosen so that $\epsilon_{0123} = \sqrt{-g}$. 
As described in the preliminary section, decompose the 
Einstein 3-form as
\begin{equation}
{\sf G}_{{A}} = {}^{\parasmall}\!{\sf G}_{{A}}
+ e^{\bot} \wedge {\sf G}_{{A}\bot} {\,} .
\end{equation}
Letting $\star$ denote the Hodge-duality defined by the 
$\Sigma$ metric $h_{ij}$, we have the relations 
${}^{\star}{\sf G}_{{A}\bot} = - G_{j{A}}e^{j}$ and 
${}^{\star}{}^{\parasmall}\!{\sf G}_{{A}} = - G_{\bot{A}}$.
When working with the $\star$ duality in index notation, the 
$\Sigma$ permutation tensor $\epsilon_{ijk}$ is the 
appropriate one to use.

Let us write down {\sc df} equations which capture 
the antisymmetric pieces of the Einstein tensor. These will 
be related to the residual torsion $T_{ij}$ in the next 
subsection. First, one immediately finds that
\begin{equation}
  G_{ij} e^{i} \wedge e^{j}
  = e^{{\rm a}} \wedge 
    {}^{\star}{\sf G}_{{\rm a}\bot} 
{\,} .
\label{antiG_ij}
\end{equation}
Now turn to the antisymmetry in the cross components 
$G_{\bot j} - G_{j\bot}$. We have already seen that
${}^{\star}{}^{\parasmall}\!{\sf G}_{{\rm a}} 
= - G_{\bot{\rm a}}$.
To get the other cross term look at ${\sf G}_{\bot\bot}$,
which obeys ${}^{\star}{\sf G}_{\bot\bot} =
- G_{j\bot}e^{j}$. Therefore, the 1-form expression
\begin{equation}
 (G_{\bot j} - G_{j\bot})e^{j}
  = - ({}^{\star} 
   {}^{\parasmall}\!{\sf G}_{{\rm a}}) e^{{\rm a}}
   + {}^{\star} {\sf G}_{\bot\bot} 
\label{antiG_iperp}
\end{equation}
captures the cross-term antisymmetry.

\subsection{Bianchi Identity for the Torsion 2-form}
 
In this subsection we shall relate the residual torsion 
$T$ to the antisymmetric pieces of the Einstein tensor. 
To do this, 
we apply the form decompositions 
(\ref{receipe},\ref{decomp}) to the Bianchi 
identity for the torsion,
\begin{equation}
{\cal D}T^{A} \equiv {\rm d}T^{A} 
+ \Gamma^{A}{}_{B} \wedge T^{B}
= \Re^{A}{}_{B} \wedge e^{B} 
{\,} , \label{Bianchitorsion}
\end{equation}
subject to the assumption that both 
$T^{{\rm a}} = 0$ and $i_{\bot} T^{\bot} = 0$. 

The $\bot$ value of (\ref{Bianchitorsion}) is then
\begin{equation}
{\rm d}T 
= \Re^{\bot}{}_{{\rm a}} \wedge e^{{\rm a}}
{\,} . \label{perptorsion}
\end{equation}
The ${}^{\parasmall}$ projection of (\ref{perptorsion}) leads
to an equation for the $\Sigma$ expression $dT$ which is of no 
concern to us here. However, taking the $i_{\bot}$ inner 
derivative of (\ref{perptorsion}) and using the form 
decompositions (\ref{receipe},\ref{decomp}), we obtain
the following identity:
\begin{equation}
N^{-1} \hat{\partial}_{0} T 
= \Re^{\bot}{}_{{\rm a}\bot{\rm b}} 
   e^{{\rm b}} \wedge e^{{\rm a}}  
{\,} .
\label{pairexchange}
\end{equation}
In order to re-express (\ref{pairexchange}) in terms of the
Einstein 3-form, first use the relations\footnote{In these 
formulae $\Re_{ij}$ represents the spatial components of the 
spacetime Ricci tensor.} 
   $\Re^{\bot}{}_{{\rm a}\bot{\rm b}}  
   = e_{{\rm a}}{}^{i}e_{{\rm b}}{}^{j}\Re_{ij} 
   - \Re^{{\rm c}}{}_{{\rm a}{\rm c}{\rm b}}$ 
and 
   $\Re_{[ij]} = G_{[ij]}$
to rewrite the preceding identity as
\begin{equation}
  G_{ij} e^{i} \wedge e^{j} =
  - N^{-1} \hat{\partial}_{0}T
  + \Re^{{\rm c}}{}_{{\rm a}{\rm c}{\rm b}}
    e^{{\rm a}} \wedge e^{{\rm b}} 
{\,} . 
\label{antiRe_ij}
\end{equation}
Next, plug the Gauss-Codazzi-Mainardi ({\sc gcm}) 
equation (\ref{triadGCM}a) 
into (\ref{antiRe_ij}), use the fact that the $\Sigma$ Ricci 
tensor $R_{ij}$ is symmetric (since we have spatial torsion 
$T^{i}{}_{jk} = 0$), and appeal to (\ref{antiG_ij}). These steps
lead to a key identity
\begin{equation}
  \hat{\partial}_{0} T 
  = N H T + N K_{{\rm a}} \wedge {}^{\star} 
   (e^{{\rm a}} \wedge {}^{\star} T) 
   - N e^{{\rm a}} \wedge {}^{\star}
    {\sf G}_{{\rm a}\bot}
{\,} ,
\label{dotT} 
\eqnum{\ref{dotT}}
\addtocounter{equation}{1}
\end{equation}
which gives the ``time derivative'' $\hat{\partial}_{0} T$ of 
the our residual torsion. This important equation establishes 
that, given the vanishing of the torsion $T|_{\Sigma}$ pulled 
back to some initial 
Cauchy surface, no torsion can develop for 
vacuum spacetimes characterized by a vanishing Einstein 
3-form.

We now turn to the triad values of the Cartan-Bianchi identity 
(\ref{Bianchitorsion}) for the torsion, which for our restricted
torsion take the following form:
\begin{equation}
  \Gamma^{{\rm a}}{}_{\bot} \wedge T 
 = \Re^{{\rm a}}{}_{B} 
  \wedge e^{{\rm B}} 
{\,} . 
\label{atorsion}
\end{equation}
Using the {\sc gcm} equation (\ref{triadGCM}a)
for the $3+1$ splitting of the triad components of the 
spacetime Riemann tensor, one finds that the ${}^{\parasmall}$ 
projection (\ref{atorsion}) is an identity which is trivially 
satisfied. However, taking the $i_{\bot}$ inner derivative of
(\ref{atorsion}), one obtains
\begin{equation}
   a^{{\rm a}} T 
   = \Re^{{\rm a}}{}_{{\rm b}\bot}
     \wedge e^{{\rm b}}
   + \Re^{{\rm a}}{}_{\bot} 
{\,} ,
\end{equation}
which after a few more manipulations yields
\begin{equation} 
   ({}^{\star}{}^{\parasmall}\!{\sf G}^{{\rm a}}) 
   e^{\star}_{{\rm a}}
   - {\sf G}_{\bot\bot} = a \wedge {}^{\star} T 
{\,} .
\label{T_is_antiG}
\end{equation}
Here and in what follows
we make extensive use of the following the common short-hands:
\begin{eqnarray}
  e^{\star}_{\rm ab} & \equiv & 
  \epsilon_{\rm abc} e^{\rm c}
\label{shorthands} \eqnum{\ref{shorthands}a} \\
  e^{\star}_{{\rm a}} & \equiv & {\textstyle \frac{1}{2}} 
  \epsilon_{{\rm a}{\rm b}{\rm c}} 
  e^{{\rm b}} \wedge e^{{\rm c}} 
{\,} .
\eqnum{\ref{shorthands}b}
\addtocounter{equation}{1}
\end{eqnarray}
One should understand that 
${}^{\star}{}^{\parasmall}\!{\sf G}^{{\rm a}}$ is a 0-form. 
Looking back at (\ref{antiG_ij}), we see that this identity
relates the torsion 2-form to the cross-term antisymmetry
of the Einstein tensor. In index notation, the equation
\begin{equation}
G_{i\bot} - G_{\bot i} = T_{ij} a^{j}
\label{indexT_is_antiG}
\end{equation}
captures this antisymmetry.


\section{Sparling forms and the $3+1$ Einstein equations}  

Sparling's differential forms have played a central role in
several recent developments in general relativity,\cite{mason} 
and the tetrad versions\cite{Madore,Goldberg} of these forms also 
play a principal role in our formalism. We begin this section by 
introducing the tetrad Sparling forms and by reviewing 
their remarkable properties. In particular, we recall that the 
Sparling forms provide a handy expression for the Einstein 
3-form. Meshing quite nicely with our decomposition 
(\ref{decomp}) of exact spacetime forms, this handy expression
leads to a $3+1$ decomposition of the Einstein 3-form which
is particularly useful in our construction of a nonlinear wave
equation for general relativity. We consider the special case 
when the (tetrad-dependent) Sparling forms are determined by a 
time-gauge tetrad (\ref{tgtetrad}). Although the results of 
this section are standard ones, it would appear that we perform 
the 3+1 decomposition in a novel way.

\subsection{Basic Properties and Time-gauge Expressions}

We start with the 
{\em Sparling relation},\cite{Madore,Goldberg} which expresses 
the Einstein 3-form
\begin{equation}
    {\sf G}_{{A}} = {\rm d}\sigma_{{A}} 
  - \tau_{{A}} - {\textstyle \frac{1}{2}} 
    \epsilon_{ABCD} T^{B} \wedge \Gamma^{CD}
\label{Sparlingrelation}
\end{equation}
in terms of the torsion 2-form, the tetrad connection 
coefficients, and the tetrad-dependent Sparling 2-forms and 
3-forms,
\begin{eqnarray}
  \sigma_{{A}} & \equiv & 
- {\textstyle \frac{1}{2}}\Gamma^{{B}{C}} 
  \wedge e^{\ast}_{{A}{B}{C}}   
\label{0sigSparling} \\
& & \nonumber \\
  \tau_{{A}} & \equiv & 
  {\textstyle \frac{1}{2}}
  \left(\Gamma_{{A}}\,^{{B}} 
  \wedge \Gamma^{{C}{D}} 
  \wedge e^{\ast}_{{B}{C}{D}} - 
  \Gamma^{{B}}\,_{{D}} 
  \wedge \Gamma^{{D}{C}} 
  \wedge e^{\ast}_{{A}{B}{C}}\right)
{\,}.
\label{0tauSparling}
\end{eqnarray}
Here 
$e^{\ast}_{{A}{B}{C}} \equiv \epsilon_{{A}{B}{C}{D}} e^{{D}}$.
Notice that, due to the way in which the tetrad connection 
coefficients appear in the definitions above, the Sparling forms 
do not behave homogeneously under tetrad transformations. 
However, while individually neither ${\rm d}\sigma_{{A}}$, 
$\tau_{{A}}$, nor the final torsional 
term in the Sparling relation 
(\ref{Sparlingrelation}) behave homogeneously, the sum of 
terms on the right-hand side of (\ref{Sparlingrelation}) must 
behave homogeneously under tetrad transformations, since this 
sum is the Einstein 3-form. 

We shall have need to consider only the Sparling forms 
$\{\sigma_{\bot},\sigma_{{\rm a}},\tau_{\bot},\tau_{{\rm a}}\}$ 
determined by a time-gauge tetrad (\ref{tgtetrad}). 
Consider first the ${}^{\parasmall}$ and ${}^{\perp}$ projections of 
the 2-forms $\{\sigma_{\bot},\sigma_{{\rm a}}\}$.
One quickly finds the following list:
\begin{eqnarray}
  {}^{\parasmall}\!\sigma_{{\rm a}} & = & 
   e^{\star}_{{\rm a}{\rm b}} \wedge
   K^{{\rm b}}
\label{sigset} \eqnum{\ref{sigset}a} \\
  \sigma_{{\rm a}\bot} & = & 
  \omega_{{\rm a}} 
  + a^{{\rm b}} e^{\star}_{\rm ab}
\eqnum{\ref{sigset}b} \\
  {}^{\parasmall}\!\sigma_{\bot} & = & 
  \omega_{{\rm a}} \wedge e^{{\rm a}}
\eqnum{\ref{sigset}c} \\
  \sigma_{\bot\bot} & = & 
  N^{-1} \Gamma_{{\rm a}0} e^{{\rm a}} 
{\,},
\eqnum{\ref{sigset}d}
\addtocounter{equation}{1}
\end{eqnarray}
where for notational convenience we define $\omega_{{\rm a}} 
\equiv - \frac{1}{2} \epsilon_{{\rm a}{\rm b}{\rm c}} 
\omega^{{\rm b}{\rm c}}$. We make the same definitions for 
$\Gamma_{{\rm a}0}$ and the triad-valued curvature two-form 
$R_{{\rm a}}$ below in (\ref{tauset}c). Notice that the forms 
${}^{\parasmall}\! \sigma_{{\rm a}}$ and 
$\sigma_{{\rm a}\bot} - \omega_{{\rm a}}$, behave homogeneously 
under triad transformations, and hence that it makes
sense to consider their exterior covariant derivatives,  
$D{}^{\parasmall}\! \sigma_{{\rm a}}$ and 
$D(\sigma_{{\rm a}\bot} - \omega_{{\rm a}})$. Evidently,
the expressions for $\sigma_{\bot\bot}$ and 
${}^{\parasmall}\! \sigma_{\bot}$ behave inhomogeneously 
under rotations of the triad. 

The task of finding the explicit time-gauge expressions for
the set $\{{}^{\parasmall}\!\tau_{{\rm a}} , \tau_{{\rm a}\bot} ,
{}^{\parasmall}\! \tau_{\bot} , \tau_{\bot\bot}\}$ is somewhat
tedious, and to get the expressions in the list below, one must 
make extensive use of $\epsilon_{{\rm a}{\rm b}{\rm c}}$ 
permutation-symbol gymnastics. We find
the following set:
\begin{eqnarray}
    {}^{\parasmall}\! \tau_{{\rm a}} & = & 
  - \omega^{{\rm b}}{}_{{\rm a}} 
    \wedge K^{{\rm c}} \wedge e^{\star}_{{\rm b}{\rm c}}
  + {\textstyle \frac{1}{2}} 
    \epsilon_{{\rm a}{\rm b}{\rm c}}
    \omega^{{\rm b}{\rm c}} \wedge T 
\label{tauset} \eqnum{\ref{tauset}a} \\
    {}^{\parasmall}\! \tau_{\bot} & = & 
    {\textstyle \frac{1}{2}}
    K^{{\rm a}} \wedge K^{{\rm b}} \wedge 
    e^{\star}_{{\rm a}{\rm b}} 
  - {\textstyle \frac{1}{2}} 
    \omega^{{\rm a}}\,_{{\rm d}} \wedge
    \omega^{{\rm d}{\rm b}} \wedge
    e^{\star}_{{\rm a}{\rm b}} 
\eqnum{\ref{tauset}b} \\
    \tau_{{\rm a}\bot} & = & 
  - N^{-1} 
    \Gamma^{{\rm b}}\,_{{\rm a}0} K^{{\rm c}} 
    \wedge e^{\star}_{{\rm b}{\rm c}} 
  + R_{{\rm a}} - d\omega_{{\rm a}} 
  - a \wedge \omega_{{\rm a}}
\nonumber \\
& & 
  + a^{{\rm b}} 
    \omega^{{\rm c}}\,_{{\rm a}} \wedge 
    e^{\star}_{{\rm b}{\rm c}}  
  + {\textstyle \frac{1}{2}}
    \epsilon_{{\rm a}{\rm b}{\rm c}} K^{{\rm b}} 
    \wedge K^{{\rm c}} - N^{-1} \Gamma_{{\rm a}0} T
\eqnum{\ref{tauset}c} \\
    \tau_{\bot\bot} & = & 
  - a^{{\rm a}} K^{{\rm b}} \wedge 
    e^{\star}_{{\rm a}{\rm b}} 
  - {\textstyle \frac{1}{2}}\epsilon_{{\rm a}{\rm b}{\rm c}} 
    K^{{\rm a}} \wedge \omega^{{\rm b}{\rm c}} 
  - N^{-1} \Gamma^{{\rm a}}\,_{{\rm d}0}
    \omega^{{\rm d}{\rm b}} \wedge 
    e^{\star}_{{\rm a}{\rm b}}
{\,} ,
\eqnum{\ref{tauset}d}  
\addtocounter{equation}{1}
\end{eqnarray}
where in (\ref{tauset}c) one finds the $\Sigma$ 
triad-valued curvature 2-form $R_{{\rm a}}
= d\omega_{{\rm a}} + \frac{1}{2} 
\epsilon_{{\rm a}{\rm b}{\rm c}} \omega^{{\rm b}} \wedge 
\omega^{{\rm c}}$.

We shall be dealing exclusively with the projected 
2-form $^{\parasmall}\!\sigma_{{\rm a}}$. 
Therefore, to avoid undue 
notational clutter, let us define the $\Sigma$ form 
\begin{equation}
\pi_{{\rm a}} \equiv - {\textstyle \frac{1}{2}}{\,} 
{}^{\parasmall}\!\sigma_{{\rm a}} = 
{\textstyle \frac{1}{2}}
K^{{\rm b}} \wedge e^{\star}_{{\rm a}{\rm b}}
{\,} 
\label{pi_a}
\end{equation}
(the factor of $-\frac{1}{2}$ has been chosen for later
convenience) and collect a few results concerning it.
First, with the definition (\ref{shorthands}b), 
the evolution rule (\ref{triadevolution}) for the 
cotriad shows that 
\begin{equation}
\hat{D}_{0} e^{\star}_{{\rm a}} = - 2N \pi_{{\rm a}}
{\,} .
\label{pidot}
\end{equation}
Moreover, defining\footnote{Note that this is a 
special definition,
and that $\pi^{{\rm a}{\rm b}}$ is not obtained 
from $\pi^{{\rm b}}$
via inner differentiation, i.~e.~$\pi_{{\rm a}{\rm b}}
\not\equiv i_{{\rm b}} \pi_{{\rm a}}$.} 
\begin{equation}
\pi^{{\rm a}{\rm b}}
\equiv e^{{\rm a}} \wedge \pi^{{\rm b}} = 
{\textstyle \frac{1}{2}}
(H \delta^{{\rm a}{\rm b}} - K^{{\rm a}{\rm b}}) e^{\star}
{\,} ,
\label{definition_pi_ab}
\end{equation}
we take the {\em rotation-constraint 3-form} to be
\begin{equation}
J^{{\rm a}{\rm b}} 
= - 4 \pi^{[{\rm a}{\rm b}]}
{\,} . 
\label{rotation3form}
\end{equation}
In fact, $J^{{\rm a}{\rm b}} = \delta^{{\rm a}{\rm c}}
\delta^{{\rm b}{\rm d}} T_{{\rm c}{\rm d}} e^{\star}$, where
the $T_{{\rm c}{\rm d}}$ are the triad components of the 
residual torsion 2-form. Alternatively, the construction 
\begin{equation}
T = 2 e^{{\rm a}} \wedge {}^{\star}\!\pi_{{\rm a}} 
{\,} 
\label{T_from_pi}
\end{equation}
provides a direct way of expressing the torsion 2-form. 

We wish to point out that 2-form $\pi_{{\rm a}}$ is 
essentially the momentum conjugate 
to the cotriad in the Hamiltonian formulation of triad 
gravity.\cite{Goldberg} Indeed, (switching to index notation) 
the $\Sigma$ Hodge dual of $\pi_{{\rm a}ij}$ is  
\begin{equation}
{}^{\star}\! \pi_{{\rm a}}{}^{k} 
= 8 \pi h^{-1/2} e_{{\rm a} i} \pi_{\scriptscriptstyle ADM}^{ki}{},
\end{equation}
where $\pi_{\scriptscriptstyle ADM}^{ij} 
= (16\pi)^{-1}\sqrt{h}(K h^{ij} - K^{ij})$ is the standard 
{\sc adm} gravitational momentum.\cite{ADM} 
Therefore, were we considering the
standard Hamiltonian triad formalism, we would have 
the following as a canonical pair: 
\begin{equation}
\left\{(8\pi)^{-1} \pi_{{\rm a}ij}{\,}, {\,}
\tilde{e}^{{\rm a}ij}{}\right\} 
{\,}, \label{newE}
\end{equation}
where $\tilde{e}^{{\rm a}ij} \equiv \sqrt{h}\epsilon^{k ij} 
e^{{\rm a}}\,_{k}$. The standard {\em rotation constraint} which
arises in Hamiltonian triad gravity 
(also known as the {\em Gauss
constraint} in the Ashtekar formulation \cite{Ashtekar}) 
is nothing but our 
$J^{{\rm a}{\rm b}}$ (stripped of a $d^{3}x$). The vanishing of
the rotation constraint is equivalent to the vanishing of our
residual torsion 2-form $T$.

\subsection{3+1 Decomposition of the Einstein 3-form}
Using the form decompositions (\ref{receipe}) and 
(\ref{decomp}) and recalling that the only non-zero piece
of the torsion 2-form is $T = {}^{\parasmall}\!T^{\bot}$, 
we immediately write the Sparling relation 
(\ref{Sparlingrelation}) as the following $\bot$ and 
${}^{\parasmall}$ projections:
\begin{eqnarray}
  {\sf G}_{{A}\bot} & = & 
  N^{-1}\left[\hat{\partial}_{0}
  {}^{\parasmall}\!\sigma_{{A}}
  - d (N \sigma_{{A}\bot})\right] 
  - \tau_{{A}\bot} 
  - \eta^{{\rm b}}_{A} N^{-1} \Gamma_{{\rm b}0} T
\nonumber \\
& &  \label{pieces} \\
  {}^{\parasmall}\!{\sf G}_{{A}} & = & 
  d {}^{\parasmall}\!\sigma_{{A}} 
  - {}^{\parasmall}\!\tau_{{A}}
  - \eta^{{\rm b}}_{A}
  T \wedge \omega_{{\rm b}}
{\,} .
\nonumber
\end{eqnarray}
Now, in fact, we know that as a whole these pieces 
must be triad-gauge covariant. Indeed, insertion of 
the explicit expressions (\ref{sigset}) and (\ref{tauset}) 
into (\ref{pieces}) yields the following nice formulae:
\begin{eqnarray}
  {}^{\parasmall}\! {\sf G}_{{\rm a}} & = & 
  - 2D \pi_{{\rm a}} 
\label{3+1etf} \eqnum{\ref{3+1etf}a} \\
  {}^{\parasmall}\! {\sf G}_{\bot} & = & 
  2 {}^{\star}\! \pi_{{\rm a}{\rm b}}
  \pi^{{\rm b}{\rm a}}
  -  H \pi^{{\rm b}}{}_{{\rm b}} 
  + e^{{\rm a}} \wedge R_{{\rm a}}
\eqnum{\ref{3+1etf}b} \\
  {\sf G}_{{\rm a}\bot} & = & 
  - 2 N^{-1}\hat{D}_{0}\pi_{{\rm a}} 
  + N^{-1} e^{\star}_{{\rm a}{\rm b}} 
    \wedge D(N a^{{\rm b}})  - R_{{\rm a}}
\nonumber \\
& & 
  + 2 {}^{\star}\! \pi_{{\rm b}{\rm c}}
  {}^{\star}\! \pi^{{\rm c}{\rm b}} 
  e^{\star}_{{\rm a}}
  - H^{2} e^{\star}_{{\rm a}}  
  - 4{}^{\star}\! \pi_{{\rm b}{\rm a}} \pi^{{\rm b}} 
  + 2 H \pi_{{\rm a}} 
{\,} ,    
\eqnum{\ref{3+1etf}c}
\addtocounter{equation}{1}
\end{eqnarray}
where $H = {}^{\star}\! \pi^{{\rm a}}{}_{{\rm a}}$ and 
$\pi^{{\rm a}}{}_{{\rm a}} = H e^{\star}$. 
Notice that (\ref{3+1etf}a) and (\ref{3+1etf}b), respectively, 
are 3-form versions of the standard momentum and Hamiltonian 
constraints.\footnote{In \cite{canonical} Isenberg and Nester
expressed the gravitational constraints in essentially this 
way. The $\pi^{a}$ defined in a footnote on page 60 of that
reference is, apart from an overall factor of $- 4$, the 
$\pi_{{\rm a}}$ that we use here. Also, their $\pi^{ab}$ differs
from our $\pi^{{\rm a}{\rm b}}$ by a factor of $-2$.}
We could effortlessly obtain the 3+1 expression for 
${\sf G}_{\bot\bot}$ with (\ref{3+1etf}a) and our previously 
derived result (\ref{T_is_antiG}).


\section{Wave Equation for General Relativity}        

Let us quickly take stock of the situation thus far. As a 
statement of the standard 3+1 version of Einstein's theory 
in vacuo, we take Eq.~(\ref{pidot}) (in practice, the 
equation of {\em  definition} for $\pi_{\rm a}$), the 
vanishing of Eq.~(\ref{3+1etf}c) (the equation of motion for 
$\pi_{\rm a}$), and the vanishing of 
Eqs.~(\ref{rotation3form}), (\ref{3+1etf}a), and (\ref{3+1etf}b) 
(rotation, momentum, and Hamiltonian constraints). 
After making a short digression in $\S$ IV.A in order to 
discuss the Laplacian and wave operators we 
shall use in our construction, we introduce in $\S$ IV.B a 
certain triad-valued 2-form $\Upsilon_{\rm a}$ whose vanishing 
will, in effect, replace the evolution equation for 
$\pi_{\rm a}$. We then perform a $3+1$ decomposition on the 2-form 
$\Upsilon_{\rm a}$. This decomposition yields a non-linear 
wave equation for $\pi_{\rm a}$ 
(assuming a harmonic time-slicing) which serves as the key 
equation in our system of {\sc df} equations, written down 
in $\S$ IV.C, which describes vacuum general relativity.

\subsection{Digression: Laplacian and Wave Operators on $\Sigma$ 
Triad-valued 2-forms}

Define a triad-covariant 
Laplacian $\Delta$ whose action on triad-valued 
$\Sigma$ 2-forms $\psi_{{\rm a}} 
= {}^{\parasmall}\!\psi_{{\rm a}}$ is
\begin{equation}
  \Delta\psi_{{\rm a}} \equiv 
  ({}^{\star}\! D {}^{\star}\! D 
- D {}^{\star}\! D {}^{\star})\psi_{{\rm a}}
{\,} .
\end{equation}
Clearly this operator reduces to the standard {\sc df} Laplacian
when acting on ordinary 2-forms.
Our chief interest is the action of the triad-covariant
Laplacian $\Delta$ on the 2-form $\pi_{{\rm a}}$. 
A short calculation shows that
\begin{equation}
  \Delta\pi_{{\rm a}} = 
  e_{{\rm a}}{}^{j} e^{{\rm b} i}
  ( \bar{\nabla}_{m}\bar{\nabla}^{m} \pi_{ij} 
- \pi_{mj} R^{m}{}_{i} 
- \pi^{mp} R_{pjim}) 
  e^{\star}_{{\rm b}} 
{\,} ,
\label{SparlingBox}
\end{equation}
where $\pi_{ij} = \frac{1}{2}(H h_{ij} - K_{ij})$ and, in accord
with the notation of {\sc aacy}, we use $\bar{\nabla}_{i}$ to denote the
torsion-free covariant derivative operator compatible with 
$h_{ij}$. Note that $\pi^{ij}$ differs from 
$\pi_{\scriptscriptstyle ADM}^{ij}$ 
by the density factor $(8\pi)^{-1}\sqrt{h}$.
Also, $R_{ij}$ is the $\Sigma$ Ricci tensor and $R_{ijkl}$ is 
the $\Sigma$ Riemann tensor. Clearly, even upon passage to the 
tensor-index formalism 
$({}^{\star}\! D {}^{\star}\! D - D {}^{\star}\! D {}^{\star})$ 
differs from the tensor 
$\bar{\nabla}^{k}\bar{\nabla}_{k}$ used in {\sc aacy}. 
However, provided that equivalence with the Einstein equations
has been shown, the extra curvature terms buried in the form 
Laplacian can be eliminated in favor of the variables of 
interest via a definite prescription.\cite{YCBandJWY,AACY,York} 
The ability to eliminate $\Sigma$ curvature terms has proven to be 
crucial in {\sc aacy}'s construction of a (gauge-covariant and causal) 
{\sc fosh} system for Einstein's theory.

Our wave operator is\footnote{We could, of course, work instead 
with the slightly different operator 
$\smash{\hat{\stackrel{}{\Box}}}' \equiv - (N^{-1}\hat{D}_{0})^{2} 
+ \Delta$, but we shall stick with (\ref{theBox}).}
\begin{equation}
\smash{\hat{\stackrel{}{\Box}}} \pi_{{\rm a}} \equiv 
[- N^{-2}(\hat{D}_{0})^{2} + \Delta] \pi_{{\rm a}}
{\,}.
\label{theBox}
\end{equation}
Writing 
$\smash{\hat{\stackrel{}{\Box}}} \pi_{{\rm a}} = 
e_{{\rm a}}{}^{j} e^{{\rm b}i} A_{ij}
e^{\star}_{{\rm b}}$, one finds that 
\begin{eqnarray}
A_{ij} & = & [- N^{-2} (\hat{\partial}_{0})^{2} 
+ \bar{\nabla}^{k}\bar{\nabla}_{k}] \pi_{ij}
\nonumber \\
& &
+ [{\rm other\,\,terms\,\, in\,\,} 
  \pi_{ij}\,\,{\rm and}\,\,
  \hat{\partial}_{0}\pi_{ij}\,\,
  ({\rm but\,\, no\,\, second\,\, 
  derivatives\,\, of}\,\,\pi_{ij})]
{\,} ,
\label{newBox}
\end{eqnarray} 
which more than hints at the fact that the tensor-index 
formulation of our construction involves a wave equation for 
$\pi_{ij}$. The tensor wave operator in (\ref{newBox}) is one 
used by {\sc aacy}. 

\subsection{Taking Derivatives of the Field Equations}

The aforementioned 2-form $\Upsilon_{\rm a}$ is defined by the 
following equation:\footnote{See the Eq.~(\ref{Upsilon_ij}) 
for the
tensor-index definition corresponding to $\Upsilon_{\rm a}$.}
\begin{equation}
2\Upsilon_{{\rm a}} \equiv
  \hat{D}_{0} {\sf G}_{{\rm a}\bot}
- N {}^{\star}\! D{}^{\star}
  {}^{\parasmall}\!{\sf G}_{{\rm a}}  
- e^{\star}_{\rm ab} \wedge
  D{}^{\star}(N {}^{\parasmall}\!{\sf G}^{\rm b}) 
+ N a^{\rm b} ({}^{\star}{}^{\parasmall}\!{\sf G}_{{\rm b}}) 
  e^{\star}_{{\rm a}} 
- D(N {\sf T}_{{\rm a}}) 
{\,} 
\label{Upsilon}
\end{equation}
(the unimportant factor of $2$ sitting before $\Upsilon_{{\rm a}}$
is chosen only for later convenience). Here ${\sf T}_{\rm a}$ 
is a certain carefully designed 1-form which is built purely 
from torsion (and spatial derivatives of torsion) whose particular 
form [given in the appendix Eq.~(\ref{torsionterms})] need not
concern us yet. Notice that 
$\Upsilon_{\rm a}$ has been constructed by taking a 
``time derivative'' of the equation of motion (\ref{3+1etf}c),
with certain spatial derivatives of the momentum and rotation 
constraints subtracted from the result. Let us briefly comment 
on the origin of the various terms 
in the definition (\ref{Upsilon}). In performing the $3+1$ 
decomposition of $\hat{D}_{0} {\sf G}_{{\rm a}\bot}$, one generates 
certain Einstein 3-form and torsion terms, in addition to other 
terms which are built with neither Einstein 3-form nor the 
residual torsion. Now, in fact, the subsequent terms in 
$\Upsilon_{{\rm a}}$ have been tailored 
to exactly cancel these generated terms. However, we note 
that our expression for $\Upsilon_{\rm a}$ is not unique, as one 
may always add terms to $\Upsilon_{\rm a}$ which are homogeneous 
in the Einstein 3-form and torsion. Indeed, the fourth 
term in (\ref{Upsilon}), linear in the undifferentiated Einstein 
3-form, may be striken from the definition. We have included 
this term in order to obtain a more pleasing result for 
$\Upsilon_{\rm a}$'s 3+1 decomposition. However, this would seem
to amount only to a matter of taste.

To obtain the 3+1 decomposition of 2-form (\ref{Upsilon}),
perform  3+1 decomposition on the very first term 
$\hat{D}_{0}{\sf G}_{{\rm a}\bot}$ in the definition. With the 
splitting result (\ref{3+1etf}c), a straightforward expansion 
gives
\begin{eqnarray}
  \hat{D}_{0}
  {\sf G}_{{\rm a}\bot} & = & 
- 2 \hat{D}_{0}(N^{-1}\hat{D}_{0}\pi_{{\rm a}})  
+ \hat{D}_{0}
  [N^{-1} e^{\star}_{\rm ab} \wedge 
  D(N a^{{\rm b}})]  
- \hat{D}_{0} R_{{\rm a}}
\nonumber \\
& & 
+ 2 (\partial_{0}H)\pi_{{\rm a}} + 2 H \hat{D}_{0}\pi_{{\rm a}}
+ \hat{D}_{0}(2 {}^{\star}\! \pi_{{\rm b}{\rm c}}
  {}^{\star}\! \pi^{{\rm c}{\rm b}} 
  e^{\star}_{{\rm a}}
- H^{2} e^{\star}_{{\rm a}}  
- 4{}^{\star}\! \pi_{{\rm b}{\rm a}} \pi^{{\rm b}})
{\,} .\label{sfGdot}
\end{eqnarray}
The second and third terms on the right-hand side of this identity
are the ones which require some work. The third term, 
$-\hat{D}_{0}R_{\rm a}$, is handled in Appendix C; therefore, focus 
attention on the second term. 
Using the evolution rule (\ref{triadevolution}) 
for the cotriad and the commutator result 
(\ref{commutatorhatD_0andD}), we write this term as follows:
\begin{eqnarray}
\hat{D}_{0}[N^{-1}e^{\star}_{\rm ab} \wedge 
D(N a^{{\rm b}})]  & = &
- Na_{\rm d}(\Re^{\rm bd}{}_{\bot} 
+ 2 a^{[{\rm b}} K^{{\rm d}]}) \wedge e^{\star}_{\rm ab}
+ \epsilon_{{\rm a}{\rm b}{\rm c}} 
  K^{{\rm b}} \wedge D(N a^{{\rm c}})
\nonumber \\
& &
- N^{-2}(\partial_{0} N) 
  e^{\star}_{\rm ab} \wedge 
  D(N a^{{\rm b}}) 
+ N^{-1} e^{\star}_{\rm ab} \wedge D 
\hat{D}_{0}(N a^{\rm b})
{\,} .
\label{4/10point}
\end{eqnarray}
At this point, we perform some permutation-symbol 
gymnastics with the first term on the right-hand side
in order to reach\footnote{We note
that there is another possible route which one could take
at this point. Namely, one could directly substitute
the appendix result 
(\ref{AppendixBidentity}) into the first term 
of (\ref{4/10point}). If one intends to follow this route, then
one should first strike the term 
$N a^{\rm b}({}^{\star}{}^{\parasmall}\!{\sf G}_{\rm b}) 
e^{\star}_{\rm a}$ from the definition (\ref{Upsilon})
of $\Upsilon_{\rm a}$.}
\begin{eqnarray}
\lefteqn{\hat{D}_{0}[N^{-1}e^{\star}_{\rm ab} \wedge 
D(N a^{{\rm b}})]  =} & \hspace{1cm} &
\nonumber \\
& &
  N a_{{\rm a}} G_{\bot}{}^{{\rm b}} 
  e^{\star}_{{\rm b}}
- {\textstyle \frac{1}{2}} 
  \epsilon_{{\rm a}{\rm b}{\rm c}} N
  (\Re^{{\rm b}{\rm c}}{}_{\bot} 
+ 2 a^{{\rm b}} K^{{\rm c}}) \wedge a
+ N a_{{\rm a}} a_{{\rm b}} 
  (H \delta^{{\rm b}{\rm c}} 
- K^{{\rm b}{\rm c}}) e^{\star}_{{\rm c}}
\nonumber \\
& &
+ \epsilon_{{\rm a}{\rm b}{\rm c}} 
  K^{{\rm b}} \wedge D(N a^{{\rm c}})
- N^{-2}(\partial_{0} N) 
  e^{\star}_{\rm ab} \wedge 
  D(N a^{{\rm b}}) 
+ N^{-1} e^{\star}_{\rm ab} \wedge 
  D\hat{D}_{0}
  (N a^{{\rm b}})
{\,} .
\label{5/10point}
\end{eqnarray}
Next, with the index evolution rule 
(\ref{indextriadevolution}) we derive the identity
\begin{equation}
  \hat{D}_{0} (N a^{{\rm b}}) 
\equiv \hat{D}_{0} ( e^{{\rm b} i} \partial_{i} N)
= D^{{\rm b}}(\partial_{0}N) 
+ N^{2} a_{{\rm d}} K^{{\rm d}{\rm b}}
\end{equation}
(on a scalar like $\partial_{0} N$, the action of 
$D^{\rm b}$ is $e^{{\rm b}i} \partial_{i}$), and then 
insert it into the last term of 
(\ref{5/10point}), expanding the resulting
expression. At the same time, we make a substitution
into the second term in (\ref{5/10point}) with 
the appendix result (\ref{Estep}). Combining the result of
these endeavors with (\ref{sfGdot}) and the appendix 
expression (\ref{mainRdot}) for $\hat{D}_{0} R_{\rm a}$, 
we readily obtain the desired expression for 
$\hat{D}_{0} {\sf G}_{{\rm a}\bot}$. Finally, all 
extrinsic curvature terms in the expression are 
re-expressed in terms of $\pi^{\rm ab}$ via 
(\ref{definition_pi_ab}).

With the splitting result for 
$\hat{D}_{0} {\sf G}_{{\rm a}\bot}$, we express 
the 3+1 decomposition of the 2-form (\ref{Upsilon}) as
follows:
\begin{equation}
\Upsilon_{{\rm a}} = N 
 \smash{\hat{\stackrel{}{\Box}}} \pi_{{\rm a}} 
+ U^{\pi}_{{\rm a}} + U^{\scriptscriptstyle T}_{{\rm a}} 
+ V_{{\rm a}}
{\,} ,
\label{Upsilonsplit}
\end{equation}
where explicitly one has the {\em source term}
\begin{eqnarray} 
U^{\rm \pi}_{{\rm a}}  & = &
  \hat{D}_{0}(   {}^{\star}\! \pi_{{\rm b}{\rm c}}
                 {}^{\star}\! \pi^{{\rm c}{\rm b}} 
                 e^{\star}_{{\rm a}}
               - {\textstyle \frac{1}{2}} 
                 H^{2} e^{\star}_{{\rm a}}
               - 2{}^{\star}\! \pi_{{\rm b}{\rm a}} 
                 \pi^{{\rm b}}   )
+  (\partial_{0}H)\pi_{{\rm a}}
- 2 N a \wedge {}^{\star}\!D{}^{\star}\!\pi_{{\rm a}}
\nonumber \\
& &
+ N a_{{\rm a}} a_{{\rm b}} {}^{\star}\!
  \pi^{{\rm b}{\rm c}}e^{\star}_{\rm c}
+  \epsilon_{{\rm a}{\rm b}{\rm c}} 
  {}^{\star}\!\pi^{{\rm b}}{}_{{\rm d}} 
  D(N a^{{\rm c}} e^{{\rm d}})
+ 2\epsilon_{{\rm a}{\rm b}{\rm c}}N^{-1} 
  D(N^{2}  
  {}^{\star}\!\pi_{{\rm d}}{}^{{\rm b}} 
  a^{({\rm c}} e^{{\rm d})}) 
\nonumber \\
& &
- N^{-1} e^{\star}_{\rm ab} \wedge 
  D(N^{2}H a^{{\rm b}})
+ N^{-2} f 
  \hat{D}_{0} \pi_{{\rm a}}
- {\textstyle \frac{1}{2}}
  N^{-2} f e^{\star}_{\rm ab} \wedge  D(N a^{{\rm b}})
{\,} , \label{U_a}
\end{eqnarray}
the {\em torsion term}
\begin{equation}
 U^{\scriptscriptstyle T}_{\rm a} = 
  {\textstyle \frac{1}{4}}\delta_{\rm ab}
  N^{2} [{}^{\star}\!
  d(N^{-1} T)] a \wedge e^{\rm b}
- {\textstyle \frac{1}{2}}\delta_{\rm ab} N^{2}
  a \wedge {}^{\star}\!D {}^{\star}
  (N^{-1} e^{\rm b} \wedge {}^{\star}T) 
+ {\textstyle \frac{1}{2}} N^{-1} D(N^{2} a_{\rm a} {}^{\star} T)
 {\,} ,
\end{equation}
and the {\em slicing term}
\begin{equation}
  V_{{\rm a}}  = 
  {\textstyle \frac{1}{2}}N^{-1} e^{\star}_{\rm ab} \wedge
  D(D^{{\rm b}} f)
{\,} .
\label{V_a}
\end{equation}
In these expression, we have defined 
$f \equiv \partial_{0} N + N^{2} H$.
Notice that all of the second-derivative terms in 
(\ref{Upsilonsplit}) involving $\pi_{\rm a}$ 
(other than those found in 
$\smash{\hat{\stackrel{}{\Box}}}\pi_{{\rm a}}$) 
have been collected together in the 
slicing term (\ref{V_a}). Also, make note of the torsion 
appearing in $U^{\scriptscriptstyle T}_{{\rm a}}$. 
Here we consider 
$T$ as a short-hand for $2e^{\rm a} \wedge 
{}^{\star}\!\pi_{\rm a}$ [cf.~Eq.~(\ref{T_from_pi})];
and, therefore, in 
effect we have the combined expression $U_{\rm a}
\equiv U^{\pi}_{\rm a} + U^{\scriptscriptstyle T}_{\rm a}$ 
as the full source term. 
We would, of course, prefer to incorporate some or all of
$U^{\scriptscriptstyle T}_{\rm a}$ 
into the definition of (\ref{Upsilon}) in order to obtain a 
more compact expression for the source term. 
Unfortunately, as we shall soon argue, it would seem that this
is not permissible, as doing so would destroy the desired 
equivalence with Einstein's theory. However, it is possible 
to re-express the combined source term in such a way that the 
residual torsion 2-form is not so manifestly apparent. 
Indeed, following the alternative route mentioned in the 
footnote before Eq.~(\ref{5/10point}), 
one finds a combined source term
which is almost entirely ``torsion-free.'' We have chosen
the route that we have, because it happens to lead to more 
aesthetically pleasing equations than does the alternative. 
That is to say, only exact exterior derivatives appear in the 
combined source term as we have written it. Following the other 
route, one finds a lot of triad covariant derivatives, $D_{\rm b}$'s, 
surfacing in the expressions. To our mind, the chosen route 
is more in spirit with the {\sc df} language intended
to be showcased here. 

\subsection{Wave-equation Formulation of 
General Relativity}

For the vanishing of $\Upsilon_{{\rm a}}$ 
to produce a good wave-equation,
the term $V_{{\rm a}}$ must be equal to a 2-form which involves
fewer than second derivatives of $H$. As described by {\sc aacy}, there 
appears to be a number of ways to do this. For our purposes it 
suffices to enforce the simple {\em harmonic slicing} 
condition,\cite{AACY}
\begin{equation}
f \equiv \partial_{0} N + N^{2} H = 0 
{\,} .
\label{Ndot}
\end{equation}
(An obvious generalization of this equation is obtained by 
assuming that $f = f(t,x^{i})$ is a prescribed well-behaved function.) 
One immediately sees that $V_{{\rm a}}$ vanishes for a harmonic 
time-slicing. The vanishing of $V_{{\rm a}}$ also removes 
the term involving triple derivatives of the lapse. Further, two 
terms in the expression (\ref{U_a}) for $U^{\pi}_{\rm a}$ 
also vanish for
harmonic time-slicing. 

As the new Einstein system of equations of motion, we now offer
\begin{eqnarray} 
 \hat{D}_{0} e^{\star}_{{\rm a}} & = &
- 2 N \pi_{{\rm a}}
\eqnum{$\dot{e}$} \\  
 \smash{\hat{\stackrel{}{\Box}}}
\pi_{{\rm a}} & = & - N^{-1} U_{{\rm a}}
\eqnum{$\ddot{\pi}$} \\
  \partial_{0} N & = & 
- N^{2} H \eqnum{$\dot{N}$}
{\,} .
\end{eqnarray}
The appropriate Cauchy data on the initial slice are 
$e^{\star}_{\rm a}$, $\pi_{\rm a}$, $N > 0$, and 
$\hat{D}_{0}\pi_{\rm a}$ subject to the following 
constraints:
\begin{eqnarray} 
    0 & = & - 4 \pi^{[{\rm a}{\rm b}]}
\eqnum{${\sf J}$} \\
    0 & = & - 2 D \pi_{{\rm a}}
\eqnum{$\vec{\sf H}$} \\
    0 & = & 
      2 {}^{\star}\! \pi_{{\rm a}{\rm b}}
      \pi^{{\rm b}{\rm a}}
    -  H \pi^{{\rm b}}{}_{{\rm b}} 
    + e^{{\rm a}} \wedge R_{{\rm a}}
\eqnum{${\sf H}$} \\
    0 & = & 
    - 2 N^{-1}\hat{D}_{0}\pi_{{\rm a}} 
    + N^{-1} e^{\star}_{{\rm a}{\rm b}} 
      \wedge D(N a^{{\rm b}})  
    - R_{{\rm a}}
    - H^{2} e^{\star}_{{\rm a}}  
    + {}^{\star}\! \pi_{{\rm b}{\rm c}}
      {}^{\star}\! \pi^{{\rm c}{\rm b}} 
      e^{\star}_{{\rm a}} 
    + 2 H \pi_{{\rm a}}   
    - 4 {}^{\star}\! \pi_{{\rm b}{\rm a}} 
    \pi^{{\rm b}}
{\,} ,
\eqnum{$\dot{\pi}$}
\end{eqnarray}
where the understanding is that these constraint equations 
are pulled back to the initial-data surface. Notice that
we are requiring the full Einstein 3-form to vanish on the initial
data surface. Both the shift vector $\beta^{j}$ and the 
skew-symmetric rotation matrix $\phi_{\rm ab}$ play a passive 
role in our formalism, as they have been absorbed into the 
operator $\hat{D}_{0}$.


\section{Verification of the original Einstein equations}

Before showing that the new system of equations is equivalent
to the original Einstein equations, we should demonstrate that
it is indeed solvable. However, for sake of brevity, we only note
here that the treatment of this issue found in the third 
reference of \cite{AACY} goes through essentially unaltered for
the case at hand. Therefore, let us turn to the proof of 
equivalence. 

\noindent \underline{\sc Claim}: 
The vanishing of the Einstein 3-form and torsion is 
equivalent to the 2-form equation
\begin{equation}
\Upsilon_{{\rm a}} = 0
{\,} ,
\label{Upsilonequation}
\eqnum{\ref{Upsilonequation}}
\addtocounter{equation}{1}
\end{equation}
when subject to the initial-value conditions
\begin{eqnarray}
{}^{\parasmall}\!{\sf G}_{\bot} |_{\Sigma} 
& = & 0
\label{BCs}  \eqnum{\ref{BCs}a} \\ 
{}^{\parasmall}\!{\sf G}_{{\rm a}} |_{\Sigma} 
& = &  0 \eqnum{\ref{BCs}b} \\
{\sf G}_{{\rm a}\bot} |_{\Sigma} & = & 0 
\eqnum{\ref{BCs}c} \\
T |_{\Sigma} & = & 0
{\,} . \eqnum{\ref{BCs}d}
\addtocounter{equation}{1}
\end{eqnarray}
Here $|_{\Sigma}$ means ``pull-back to the initial-data
surface." 

\subsection{Method of Proof and Two Lemmas}
To prove the claim, it proves most convenient to switch
to and proceed in index notation. Our proof follows the 
arguments of {\sc aacy} and of York\cite{York} quite closely. 
We shall have need of the spacetime covariant derivative 
which ``sees'' quasi-coordinate indices, and we use 
$\nabla_{\mu}$ to denote this covariant derivative. 
Consider the action of $\hat{\partial}_{0}$ and 
$\bar{\nabla}_{i}$ on $\Sigma$ scalars (such as 
$G_{00}$), $\Sigma$ 1-forms (such as $G_{0i}$), and 
$\Sigma$ 2-index tensors (such as $G_{ij}$) which arise from 
the various components of a 2-index spacetime tensor 
(such as $G_{\alpha\beta}$). The proof is facilitated 
by relating the action of $\hat{\partial}_{0}$ and 
$\bar{\nabla}_{i}$ on these $\Sigma$ objects to the 
action of $\nabla_{0}$ and $\nabla_{i}$ on the same 
objects. Such relations follow from the definition of 
these operators and the values of the connection 
coefficients associated with the quasi-coordinate 
frame (\ref{qcframe}), those being given by
\begin{equation}
\Gamma^{\sigma}{}_{\lambda\mu}
= {\textstyle \frac{1}{2}}g^{\sigma\kappa}
  (e_{\lambda}[g_{\kappa\mu}] + 
   e_{\mu}[g_{\kappa\lambda}] - 
   e_{\kappa}[g_{\lambda\mu}])
+ {\textstyle \frac{1}{2}}(g^{\sigma\kappa}g_{\mu\nu}
  C_{\kappa\lambda}{}^{\nu} + g^{\sigma\kappa} 
  g_{\lambda\nu} C_{\kappa\mu}{}^{\nu} - 
C_{\lambda\mu}{}^{\sigma})
{\,} .
\end{equation}
Here the 
$C_{\mu\nu}{}^{\lambda} \equiv [e_{\mu}, e_{\nu}]^{\lambda}$ 
are the structure functions associated with the frame 
(\ref{qcframe}). The frame choice 
(\ref{qcframe}) ensures that the $\bar{\Gamma}{}^{i}{}_{jk} 
\equiv \Gamma^{i}{}_{jk}$ are the ordinary Christoffel 
symbols associated with the $\Sigma$ coordinate frame.
Throughout the proof, we shall employ the short-hand
{\rm (l.h.E.T.)}$_{n}$ to mean 
``additive remainder terms {\em linear} and 
{\em homogeneous} in both 
the {\em Einstein} tensor and the {\em torsion} tensor and their 
derivatives of order $\leq n$.'' When remainder
terms are built solely with the Einstein tensor or solely 
with the torsion tensor, we shall employ the short-hands 
${\rm (l.h.E.)}_{n}$ and ${\rm (l.h.T.)}_{n}$, whose meanings
should be clear. 

We begin by defining a $\Sigma$ tensor $\Upsilon_{ij}$ 
via $\Upsilon_{\rm a} = e_{\rm a}{}^{j} e^{{\rm b}i}
\Upsilon_{ij} e^{\star}_{\rm b}$. Provided that the triad is 
nowhere degenerate, the vanishing of $\Upsilon_{ij}$ $\iff$
the vanishing of $\Upsilon_{\rm a}$. Direct calculation
now yields
\begin{equation}
\Upsilon_{ij} = - \hat{\partial}_{0} G_{ij} + 
\bar{\nabla}_{i} G_{0j} + \bar{\nabla}_{j} G_{0i} - 
h_{ij} \bar{\nabla}^{l}G_{0l} - C_{ij} + 
{\rm (l.h.E.)}_{0}
{\,} .
\label{Upsilon_ij}
\end{equation}
 Hence, in (\ref{Upsilon_ij}) no derivatives of 
the Einstein tensor appear other than those explicitly shown.
The $\Sigma$ tensor $C_{ij}$ is defined via 
$e_{\rm a}{}^{j} e^{{\rm b}i} C_{ij} e^{\star}_{\rm b} 
= D(N {\sf T}_{\rm a})$, and for it we find the following
explicit expression 
[cf.~the footnote before (\ref{torsionterms})]:
\begin{equation}
C_{ij} = {\textstyle \frac{1}{2}}
         \bar{\nabla}^{l}\left\{N^{2}\left[
         \bar{\nabla}_{i} (N^{-1} T_{jl})
         + \bar{\nabla}_{j} (N^{-1} T_{il})
         + \bar{\nabla}_{l}(N^{-1} T_{ij})\right]\right\}
{\,} .  
\label{C_ij}
\end{equation}
The $i \leftrightarrow j$ symmetry present
in the first two terms proves to be quite important for our
construction. In fact, this symmetry (in tandem with the way the 
particular powers of the lapse enter the expression), ensures 
that the antisymmetric piece of $N^{-1} C_{ij}$ is the 
Laplacian on $T_{ij}$, up to a trivial ${\rm (l.h.T.)}_{0}$ 
term. Compare the definition (\ref{Upsilon_ij}) of 
$\Upsilon_{ij}$ with {\sc aacy}'s definition 
of $\Omega_{ij}$ (or, even better, with their $\Omega'_{ij}$). 
In this comparison, be sure to note that both 
$\hat{\partial}_{0} G_{ij}$ 
and $C_{ij}$ are not manifestly symmetric in our formalism.

With the aforementioned relations between the actions
of $\hat{\partial}_{0}$ and $\bar{\nabla}_{i}$ and of 
$\nabla_{0}$ and $\nabla_{i}$, we write
\begin{equation}
\Upsilon_{ij} = - \nabla_{0} G_{ij} + \nabla_{i} G_{0j}
+ \nabla_{j} G_{0i} - h_{ij} \nabla^{k} G_{0k} - C_{ij}
+ {\rm (l.h.E.)}_{0}{\,} ;
\label{Upsilon_ij(2)}
\end{equation}
and, moreover, we have the following two lemmas for 
$C_{ij}$.

\noindent \underline{\sc Lemma 1:} The expression for $C_{ij}$
may be written as
\begin{equation}
C_{ij} =   {\textstyle \frac{1}{2}} N 
           [\nabla^{l}\nabla_{i} T_{jl}
         +  \nabla^{l}\nabla_{j} T_{il}
         +  \nabla^{l}\nabla_{l} T_{ij}] 
         + \lambda_{ij}
         + {\rm (l.h.E.T.)}_{0}
{\,} ,
\label{C'_ij}
\end{equation}
where $\lambda_{ij}$ is a {\em symmetric} remainder term of 
${\rm (l.h.T.)}_{1}$ order.

To prove the lemma simply expand the expression 
(\ref{C_ij}) for $C_{ij}$. Then the proof reduces to
verifying that $\nabla_{k} \nabla_{l} T_{ij} = \bar{\nabla}_{k} 
\bar{\nabla}_{l} T_{ij} + {\rm (l.h.E.T.)}_{0}$. Now, viewed as 
a spacetime tensor, the torsion $T_{\alpha\beta}$ is tangential 
to the $\Sigma$ slices ($T_{0i} = T_{00} = 0$). It follows 
that $\nabla_{k} T_{ij} = \bar{\nabla}_{k} T_{ij}$; and, hence, 
the only troublesome terms which arise in obtaining 
the verification above are those of the form $\nabla_{0} T_{ij}$. 
But such terms may be eliminated in favor of 
${\rm (l.h.E.T.)}_{0}$ ones via (\ref{dotT}) 
[technically, via (\ref{dotT}) as we write it later 
in (\ref{bianchies}c)]. 

\noindent \underline{\sc Lemma 2:} The tensor $C_{ij}$ obeys
the property $\nabla^{i} C_{ij} = {\rm (l.h.T.)}_{1}$.

Viewed as a spacetime tensor, $C_{\alpha\beta}$ is 
by definition purely spatial ($C_{0i} = C_{i0} = 
C_{00} = 0$); it follows that $\nabla^{i} C_{ij} = 
\bar{\nabla}^{i} C_{ij}$. With this fact, one can prove
the lemma directly by forming $\bar{\nabla}^{i} C_{ij}$ 
and making appeals to the Ricci identity. 
Alternatively, with the definition of $C_{ij}$ 
one may show that
\begin{equation}
e_{\rm a}{}^{j} \bar{\nabla}^{i} C_{ij} = 
{}^{\star}\!D^{2}(N {\sf T}_{\rm a})
{\,} .
\end{equation} 
Therefore, since 
${\sf T}_{\rm a}$ is ${\rm (l.h.T.)}_{1}$ by construction,
the Cartan identity for the creation of curvature 
(symbolically $D^{2} = R \wedge$) yields the result 
(provided a non-degenerate triad).

\subsection{Zero-order Wave Equation for the Torsion}

Before turning to the main part of the proof, let us
derive a 
certain zero-order wave equation for the residual 
torsion tensor $T_{ij}$ which is implicitly defined by 
the vanishing of $\Upsilon_{ij}$. This derivation 
will also further elucidate the precise role played 
in our formalism by the 1-form ${\sf T}_{\rm a}$ 
appearing in the definition (\ref{Upsilon}).
The form of (\ref{C'_ij}) ensures
that the antisymmetric piece of $C_{ij}$ obeys
\begin{equation}
C_{[ij]} = {\textstyle \frac{1}{2}} N 
\nabla^{l}\nabla_{l} T_{ij} + {\rm (l.h.E.T.)}_{0}
{\,} , \label{Laplacian_on_T}
\end{equation}
Note that the extra terms in this equation are linear 
and homogeneous in the Einstein and torsion tensors
with {\em no} spatial derivatives of them appearing. 
In particular, the tensor $\lambda_{ij}$ has been
killed by the antisymmetrization. 

Now consider the index version of (\ref{dotT}),
which may be expressed as
\begin{equation}
\nabla_{0} T_{ij} - 2N
G_{[ij]} = {\rm (l.h.T.)}_{0} {\,}
{\,} . 
\label{indexdotT}
\end{equation}
With 
Eq.~(\ref{Laplacian_on_T}) in mind, we apply $\nabla^{0}$
[the same as applying $N^{-2} \nabla_{0}$ for our frame
(\ref{qcframe})] to equation (\ref{indexdotT}) and expand. 
Along the way, we make further substitutions with 
(\ref{indexdotT}) to replace $\nabla_{0} {\rm (l.h.T.)}_{0}$
terms with ${\rm (l.h.E.T.)}_{0}$ ones. We then reach
\begin{equation}
\nabla^{0} \nabla_{0} T_{ij} - 2 N^{-1} 
\nabla_{0} G_{[ij]} = 
{\rm (l.h.E.T.)}_{0}
{\,} .
\label{middlestep}
\end{equation}
Next, substituting 
(\ref{Upsilon_ij(2)}) into (\ref{middlestep}) and appealing 
to (\ref{Laplacian_on_T}), we obtain
\begin{equation}
\Box T_{ij} = - 2 N^{-1} \Upsilon_{[ij]} 
+ {\rm (l.h.E.T.)}_{0}
{\,} ,
\label{Box_T}
\end{equation}
with $\Box$ denoting the standard covariant D' Alembertian
$\nabla_{\mu} \nabla^{\mu}$. Eq.~(\ref{Box_T}) shows 
that the vanishing of $\Upsilon_{ij}$ yields a {\em zero}-order 
wave equation for the residual torsion 2-form. This zero-order 
character proves to be quintessential in establishing 
equivalence between our {\sc df} system of equations given above and 
the vacuum Einstein equations. In fact, we have rigged the 
definition [cf.~Eqs.~(\ref{Dstep}) and 
(\ref{torsionterms})] of ${\sf T}_{\rm a}$ in (\ref{Upsilon}) 
to ensure this zero-order character for (\ref{Box_T}). 
Note that there is a projection
into $\Sigma$ implicit in (\ref{Box_T}). That is to say, while 
$T_{0i} = 0$ 
for the torsion viewed as a spacetime tensor $T_{\alpha\beta}$,
it is not the case that $\Box T_{0i}$, the tangential-normal
piece of $\Box T_{\alpha\beta}$, need be zero (of course,
$\Box T_{00}$ does indeed vanish by antisymmetry). Eq.~(\ref{Box_T})
has nothing to say about $\Box T_{0i}$; however, for our
purposes this observation is inconsequential.

\subsection{Twice-contracted Bianchi Identities and Equivalence
Proof}

We shall require the so-called twice-contracted Bianchi identities 
$\nabla^{\mu} G_{\mu\nu} = 0$ (index ordering is important!),
which we expand out,
\begin{eqnarray}
0 
& = & \nabla^{0}G_{00} + \nabla^{j} G_{j0}
\label{bianchies}\eqnum{\ref{bianchies}a}\\
0 & = & \nabla^{0} G_{0i} + \nabla^{j} G_{ji}
{\,} .
\eqnum{\ref{bianchies}b}
\addtocounter{equation}{1}
\end{eqnarray}
As is well-known, these identities are 
related to the preservation in time 
of the Hamiltonian and momentum constraints. Likewise, 
our identity (\ref{indexdotT}), which we conveniently 
write now as 
\begin{equation}
0 = \nabla_{0} T_{ij} + {\rm (l.h.E.T.)}_{0}
\eqnum{\ref{bianchies}c}
{\,} ,
\end{equation}
is derived from the Bianchi identity for the torsion
and is related to the preservation in time of the 
rotation constraint.  

As the first step in the proof, 
form $\nabla^{i} \Upsilon_{ij}$, use the Ricci identity to 
commute covariant derivatives, insert (\ref{bianchies}b),
and appeal to the second lemma, 
in order to obtain
\begin{equation}
\Box G_{0j} =
{\rm (l.h.E.T.)}_{1}
{\,} .
\label{BoxG_0j}
\end{equation}
As the second step, apply $\Box$ to the identity 
(\ref{bianchies}a). Repeated use of the Ricci 
identity on the
resulting expression yields 
\begin{equation}
\nabla^{0} \Box G_{00} + \nabla^{j} \Box G_{j0} =
{\rm (l.h.E.)}_{1}
{\,} .
\end{equation}
At this point, we insert the 
result $G_{j0} = G_{0j} + T_{jk} 
\nabla^{k} N$ [cf.~Eq.~(\ref{indexT_is_antiG})]. 
Then, appeals
to the earlier findings (\ref{Box_T}) and 
(\ref{BoxG_0j}) show that
\begin{equation}
\nabla_{0}\Box G_{00} = {\rm (l.h.E.T)}_{2}
{\,} .
\label{dotBoxG_00}
\end{equation}
Finally, as the proof's third step, apply $\Box$ to the 
definition of $N^{-1}\Upsilon_{ij}$ (which vanishes) as 
given in (\ref{Upsilon_ij(2)}), expand out the resulting 
expression, repeatedly use the Ricci identity, and make 
several appeals to (\ref{BoxG_0j}), thereby arriving
at
\begin{equation}
N^{-1} \nabla_{0} \Box G_{ij} + \Box( N^{-1} C_{ij}) =
{\rm (l.h.E.T.)}_{2}
{\,}.
\end{equation}
The lapse factor has been included to remove the lapse term
in $C_{ij}$ [cf.~the form of this tensor given in
the first lemma, Eq.~(\ref{C'_ij})]. Note that $\Box 
(N^{-1} C_{ij})$ is ${\rm (l.h.T.)}_{4}$; however, that it 
is also ${\rm (l.h.E.T.)}_{2}$ may be shown by the 
following argument which makes use of (\ref{C'_ij}). For the
leading terms in $N^{-1} C_{ij}$, those quadratic in the
covariant derivatives, we can move the action of the 
$\Box$ onto the torsion tensors in the expression. In so 
moving the $\Box$, we must again repeatedly use the 
Ricci identity, a process which generates only byproduct 
${\rm (l.h.T.)}_{2}$ terms. For $\Box (N^{-1} \lambda_{ij})$, 
{\em a priori} ${\rm (l.h.T.)}_{3}$, we note that, again up 
to ${\rm (l.h.T.)}_{2}$ terms, the $\Box$ can be 
moved onto the torsion tensors in the expression. 
Therefore,  invoking (\ref{Box_T}), we get
\begin{equation}
\nabla_{0} \Box G_{ij} = {\rm (l.h.E.T.)}_{2}
{\,} .
\label{dotBoxG_ij}
\end{equation}
Notice that the application of $\Box$ has mapped 
${\rm (l.h.T)}_{2}$ terms into ${\rm (l.h.E.T)}_{2}$ ones.
The ability to take this crucial step rests squarely on
the {\em zero}-order character of the wave equation 
(\ref{Box_T}). 

With (\ref{BoxG_0j}), (\ref{dotBoxG_00}), (\ref{dotBoxG_ij}),
we obtain the 
following system of equations:
\begin{eqnarray}
\hat{\partial}_{0} \Box G_{0j} & = & {\rm (l.h.E.T.)}_{2}
\label{formsystem} \eqnum{\ref{formsystem}a}\\
\hat{\partial}_{0}\Box G_{00} & = &
{\rm (l.h.E.T.)}_{2}
\eqnum{\ref{formsystem}b} \\
\hat{\partial}_{0} \Box G_{ij} & = &
{\rm (l.h.E.T.)}_{2} \eqnum{\ref{formsystem}c} \\
\hat{\partial}_{0} T_{ij} & = & {\rm (l.h.E.T.)}_{0}
\eqnum{\ref{formsystem}d}
{\,} .
\addtocounter{equation}{1}
\end{eqnarray}
This  system is (up to homogeneous Einstein-tensor and 
torsion-tensor terms) essentially the system 
considered in the works of {\sc aacy}, although here we see
the novel appearance of the torsion tensor. It is a 
strictly hyperbolic system. All {\rm (l.h.E.T.)}$_{2}$ and
{\rm (l.h.E.T.)}$_{0}$ terms on the right-hand side 
may be shown to be zero on the initial Cauchy 
surface by (i) the (index versions of the) boundary 
conditions (\ref{BCs}),
(ii) the Bianchi identities (\ref{bianchies}), 
(iii) the vanishing of $\Upsilon_{ij}$, and (iv) 
derivatives of the previous three items. We infer that 
$T_{ij}$ and all pieces of the Einstein tensor save 
$G_{j0}$ vanish in the future domain of dependence
$D^{+}(\Sigma)$. We then get the vanishing of $G_{j0}$ 
from the vanishing of $G_{0j}$ and $T_{ij}$ by the 
identity (\ref{indexT_is_antiG}). Equivalence has thus
been established.


\section{Conclusion}

We conclude by giving some basic instructions (with
cautionary remarks) for rewriting our {\sc df} 
results in the tensor-index language. At the outset, 
we must admit that we find passing back and forth between 
the {\sc df} and tensor-index versions of 
wave-equation general relativity 
to be not as clean and direct as we had originally 
hoped would be the case. 
Passage to the tensor-index version of the formalism 
essentially amounts to (i) defining $\Upsilon_{ij}$
as before in Eq.~(\ref{Upsilon_ij}) and then (ii) 
simply setting the torsion to zero by hand in all 
expressions [hence $C_{ij} = 0$ in (\ref{Upsilon_ij})], 
thereby obtaining a symmetric $\Upsilon_{ij}$.
Now, it is tempting to simply use $\Upsilon_{ij}$ as the 
jumping-off point for deriving an tensor-index wave 
equation for $\pi_{ij}$, working with $\Upsilon_{ij}$ 
in the same fashion that {\sc aacy} work with their 
$\Omega_{ij}$. However, as we shall now argue, 
though this is essentially the correct procedure, 
one must be careful to extract superfluous Einstein 
tensor terms which are indeed present in $\Upsilon_{ij}$. 
In (\ref{Upsilon_ij}) these superfluous terms have simply 
been swept into the ${\rm (l.h.E.)}_{0}$ short-hand,
but let us now examine the origin and nature of these
terms in more detail.

Consider again the source 
term (\ref{U_a}) written as follows: $U^{\pi}_{{\rm a}} 
= e_{{\rm a}}{}^{j} e^{{\rm b}i} U^{\pi}_{ij} e^{\star}_{{\rm b}}$.
Now, in fact, $U^{\pi}_{{\rm a}}$ still has some Einstein 
tensor ``buried in it,'' i.~e.~one 
finds that $U^{\pi}_{ij}$ contains contracted covariant
derivatives of $\pi_{ij}$, e.\ g.\ $\bar{\nabla}_{l}\pi^{l}{}_{i}$,
coming from the sixth term in the expression for $U^{\pi}_{{\rm a}}$.
But $2 \bar{\nabla}_{l}\pi^{l}{}_{i} = G_{\bot i}$, and, hence, 
in the 3+1 decomposition of $\Upsilon_{ij}$ superfluous terms 
involving the momentum constraint $G_{\bot i}$ 
arise. Therefore, one should 
locate and remove these superfluous momentum-constraint terms 
in the passage to the tensor-index formalism.
However, besides such superfluous momentum-constraint 
terms, another superfluous Einstein-tensor term also arises 
in the passage to the tensor-index version of the formalism. 
Indeed, consider the leading term in $\Upsilon_{{\rm a}}$, namely,
\begin{equation}
\hat{D}_{0} {\sf G}_{{\rm a}\bot} = 
- e_{{\rm a}}{}^{j} e^{{\rm b}i}(\hat{\partial}_{0} G_{ij}) 
  e^{\star}_{{\rm b}} - G_{ij} \hat{D}_{0}(e_{{\rm a}}{}^{j} 
  e^{{\rm b}i} e^{\star}_{{\rm b}})
{\, } .
\label{Gexpansion}
\end{equation}
The expansion (\ref{Gexpansion}) shows that our $\Sigma$ tensor
$\Upsilon_{ij}$ has  
an undifferentiated $G_{ij}$ term ``buried'' in it; hence, 
in passing to the tensor-index language
we be should be careful to locate and remove this term. 

One might ask, why not simply work with a modified definition
of $\Upsilon_{{\rm a}}$ (call it $\Upsilon_{{\rm a}}'$) such that 
passage to the tensor-index formalism is rather 
more direct, i.~e.~no Einstein tensors surface in the $3+1$ 
decomposition of $\Upsilon'_{ij}$ (obtained from $\Upsilon'_{{\rm a}} 
= e_{{\rm a}}{}^{j} e^{{\rm b}i} \Upsilon'_{ij} e^{\star}_{{\rm b}}$). 
However, the {\em differential-form} wave equation for $\pi_{{\rm a}}$ 
obtained from the corresponding 2-form equation $(\Upsilon')$ is quite 
a bit messier than the one we have considered, and, if fact, this 
wave equation contains superfluous Einstein 3-form terms in it! 
Therefore, if the intent is to remain within the 
framework of {\sc df}, then we find it preferable 
to work with (\ref{Upsilonequation}) 
and its 3+1 decomposition as given. However,
this means that one must not be too naive when rewriting the results
in the tensor-index language. One must be careful to extract
superfluous Einstein-tensor terms which will arise in the tensor 
equations directly obtained from the corresponding {\sc df} 
equations. The non-triviality associated with going back and forth 
between the two languages can be traced to whether it is the Einstein 
3-form or the Einstein tensor which is held as the 
fundamental object (i.~e.~which is derived from which).


\section{Acknowledgments}
Many thanks to both A.~Anderson and J.~W.~York for 
guidance concerning several aspects of this work. J.~W.~York
provided me with an early copy of his personal notes on the 
original (tensor-index) hyperbolic formulation of general 
relativity. These notes proved quintessential in the development 
of the ideas presented here. I am also grateful for comments by 
and discussions with H.~Balasin and R.~Beig.
This research has been supported by the ``Fonds 
zur F\"{o}r\-der\-ung der wis\-sen\-schaft\-lich\-en 
For\-schung'' in Austria (Lise Meitner Fellowship M-00182-PHY
and FWF Project 10.221-PHY).

\appendix

\section{Gauss-Codazzi-Mainardi Equations with Torsion}                     

The {\sc gcm} equations (see for example 
\cite{canonical,sources}) are integrability criteria which relate 
the ${\cal M}$ Riemann tensor 
$\Re^{\alpha}{}_{\beta\mu\nu}$ to 
the $\Sigma$ Riemann tensor $R^{i}{}_{jkl}$, the $\Sigma$
extrinsic curvature tensor $K_{ij}$, and other geometric
objects associated with the $\Sigma$ foliation. The form of 
these equations is complicated quite a bit by the presence of 
spacetime torsion. We shall present equations valid only for the 
limited type of spacetime torsion considered in this work. 

To derive the needed equations, one may proceed in a number of 
ways. Perhaps the best method is to apply the form decompositions 
(\ref{receipe}) and (\ref{decomp}) to the second 
Cartan structure equation for $\Re^{A}{}_{B}$,
the curvature 2-form. This straightforward 
method leads to the following expressions for the 
time-gauge components of the
${\cal M}$ Riemann tensor:
\begin{eqnarray}
  \Re_{{\rm a}{\rm b}{\rm c}{\rm d}} & = &
  R_{{\rm a}{\rm b}{\rm c}{\rm d}} 
  + K_{{\rm a}{\rm c}} K_{{\rm b}{\rm d}} 
  - K_{{\rm a}{\rm d}} K_{{\rm b}{\rm c}}
\eqnum{\ref{triadGCM}a} \\
  \Re_{\bot{\rm a}{\rm b}{\rm c}} & = & 
   D_{{\rm b}} K_{{\rm a}{\rm c}} 
   - D_{{\rm c}} K_{{\rm a}{\rm b}}
\eqnum{\ref{triadGCM}b} \\
  \Re_{{\rm a}{\rm b}\bot{\rm c}} & = & 
  N^{-1} e_{{\rm c}}{}^{j}\hat{\partial}_{0}
  \omega_{{\rm a}{\rm b}j}
  - a_{{\rm a}} K_{{\rm b}{\rm c}} 
  + a_{{\rm b}} K_{{\rm a}{\rm c}} 
  - N^{-1} \left(e_{{\rm c}}[\Gamma_{{\rm a}{\rm b}0}]
  -\Gamma_{{\rm d}{\rm b}0} 
   \omega^{{\rm d}}\,_{{\rm  a}{\rm c}} 
  -\Gamma_{{\rm a}{\rm d}0} 
   \omega^{{\rm d}}\,_{{\rm b}{\rm c}}\right)
\eqnum{\ref{triadGCM}c} \\ 
  \Re_{{\rm a}\bot{\rm b}\bot} & = &
   N^{-1} e_{{\rm b}}{}^{j} 
   \hat{D}_{0} K_{{\rm a} j}
   + D_{{\rm b}} a_{{\rm a}} +
   a_{{\rm a}} a_{{\rm b}}
{\,} .
\label{triadGCM} 
\eqnum{\ref{triadGCM}d} 
\addtocounter{equation}{1}
\end{eqnarray}

Let us express the identities (\ref{triadGCM}) in terms
of $\Sigma$ coordinate indices. Since all of the terms 
in (\ref{triadGCM}a,b) are tensorial, conversion of these
equations into the required forms is immediate. Moreover, with 
the evolution rule (\ref{indextriadevolution}) for the triad,
conversion of the final identity (\ref{triadGCM}d) into 
$\Sigma$ coordinate indices is also not difficult. However,
finding a suitable expression for (\ref{triadGCM}c) is 
more difficult. One can use the $({\cal D} T^{{\rm a}})_{\bot}$ 
piece (\ref{atorsion}) of the Bianchi identity 
(\ref{Bianchitorsion}) for the torsion 2-form, 
in order to express $\Re_{{\rm a}{\rm b}\bot{\rm c}}$ as 
$\Re_{\bot{\rm c}{\rm a}{\rm b}}$ plus terms built from the 
torsion tensor $T_{{\rm a}{\rm b}}$ and the lapse function $N$ 
(recall that without torsion the Riemann tensor is symmetric under 
such pair exchange). 
We obtain the following list:
\begin{eqnarray}
  \Re_{ijkl} & = &
   R_{ijkl} 
   + K_{ik} K_{jl} 
   - K_{il} K_{jk} 
\label{GCMe's} 
\eqnum{\ref{GCMe's}a} \\
  \Re_{\bot ijk} & = & 
  \bar{\nabla}_{j} K_{ik} 
  - \bar{\nabla}_{k} K_{ij}
\eqnum{\ref{GCMe's}b} \\
  \Re_{jk\bot i} & = &   
  \bar{\nabla}_{j} K_{ik} -
  \bar{\nabla}_{k} K_{i j} 
  + {\textstyle \frac{1}{2}} N^{-1}
  \bar{\nabla}_{i} (N T_{jk}) 
  + {\textstyle \frac{1}{2}} N \bar{\nabla}_{j} (N^{-1} T_{ki})
  - {\textstyle \frac{1}{2}} N \bar{\nabla}_{k} (N^{-1} T_{ji})
\eqnum{\ref{GCMe's}c} \\
  \Re_{i\bot j \bot} & = &  
  N^{-1} \hat{\partial}_{0}  
  K_{ij} + K^{l}{}_{i} K_{lj}
  + \bar{\nabla}_{j} a_{i} 
  + a_{i} a_{j} 
{\,} ,
\eqnum{\ref{GCMe's}d}
\addtocounter{equation}{1}
\end{eqnarray}
where we remind the reader that $T_{ij} = 2K_{[ij]}$,
and $\bar{\nabla}_{i}$ stands for the ordinary $\Sigma$ covariant
derivative.
Clearly, equations (\ref{GCMe's}b,c) exhibit the loss, due
to torsion, of the pair-exchange symmetry enjoyed by Riemann 
curvature in the absence of torsion. However, note
that (\ref{GCMe's}d) is also not symmetric under
$i \leftrightarrow j$ exchange. However, all of the terms
on the right-hand side of (\ref{GCMe's}d),
save the very first, are symmetric under 
$i \leftrightarrow j$ exchange. In particular, note that the 
term $\bar{\nabla}_{j} a_{i} = \bar{\nabla}_{j}\bar{\nabla}_{i}\log N$
is indeed symmetric, as we have set the $\Sigma$ torsion
tensor $T^{i}{}_{jk}$ manifestly equal to zero. 
The behavior of $\Re_{i\bot j\bot} = 
\Re_{\bot i \bot j}$ under the pair exchange $\bot i
\leftrightarrow \bot j$ is also evident from 
(\ref{pairexchange}).


\section{Commutator of $\hat{D}_{0}$ and $D$}

To evaluate the commutator $[\hat{D}_{0}, D]\psi^{{\rm a}}$, 
where $\psi^{{\rm a}} = {}^{\parasmall}\!\psi^{{\rm a}}$
is an arbitrary triad-valued spatial form, it suffices to 
consider the action of the squared spacetime exterior 
covariant derivative ${\cal D}{}^{2}$ on an 
arbitrary tetrad-valued spatial form $\psi^{{A}} = 
{}^{\parasmall}\!\psi^{{A}}$. Applying the form 
decompositions (\ref{receipe},\ref{decomp}) to the expression 
${\cal D} \psi^{{A}}$, one finds that
\begin{eqnarray}
^{\parasmall}\! 
({\cal D} {}\psi^{\bot})
& = & d \psi^{\bot} 
- K_{{\rm b}} \wedge \psi^{{\rm b}} 
\label{vs} \eqnum{\ref{vs}a} \\
({\cal D} {}\psi^{\bot})_{\bot} & = & 
N^{-1} \hat{\partial}_{0}\psi^{\bot} 
+ a_{{\rm b}}\psi^{{\rm b}} \eqnum{\ref{vs}b} \\
^{\parasmall}\! ({\cal D}\psi^{{\rm a}})& = &
D {}\psi^{{\rm a}} 
- K^{{\rm a}} \wedge \psi^{\bot}  \eqnum{\ref{vs}c}  \\
({\cal D} \psi^{{\rm a}})_{\bot} & = &
N^{-1} \hat{D}_{0} \psi^{{\rm a}}  
+ a^{{\rm a}}\psi^{\bot} \eqnum{\ref{vs}d}\, .
\addtocounter{equation}{1}
\end{eqnarray}
Now, a straightforward
expansion yields that
\begin{eqnarray}
{\cal D}{}^{2}\psi^{{\rm a}} 
& = & D {}^{\parasmall}\! 
({\cal D} \psi^{{\rm a}})
- K^{{\rm a}} \wedge  {}^{\parasmall}\! 
({\cal D} \psi^{\bot}) \nonumber \\
& & + e^{\bot} \wedge \left\{ N^{-1} [\hat{D}_{0} 
{}^{\parasmall}\!
({\cal D} \psi^{{\rm a}}) - D N 
({\cal D} \psi^{{\rm a}})_{\bot}] +
a^{{\rm a}}  {}^{\parasmall}\!
({\cal D} \psi^{\bot}) + K^{{\rm a}} \wedge
({\cal D} \psi^{\bot})_{\bot}\right\} {\,} .
\label{firstDsquared}\end{eqnarray}
However, by the second Cartan 
structure equation we know that the above expression is also
${\cal D}^{2}\psi^{{\rm a}} 
= \Re^{{\rm a}}\,_{{A}} \wedge \psi^{{A}}$,
where as before $\Re^{{A}}\,_{{B}}$ is the matrix-valued
curvature two-form of Cartan, which in index notation is 
represented by the mixed-components $\Re^{{A}}{}_{{B}\mu\nu}$ of
the spacetime Riemann tensor. 
Therefore, taking the $i_{\bot}$ inner derivative of 
(\ref{firstDsquared}), inserting the relations (\ref{vs}), and
collecting like terms in $\psi^{\bot}$ and 
$\psi^{{\rm a}}$, we determine the following identity:
\begin{eqnarray}
\lefteqn{\left(\Re^{{\rm a}}\,_{\bot\bot} 
+ N^{-1} \hat{D}_{0} K^{{\rm a}} + D 
a^{{\rm a}} + a^{{\rm a}} a\right) 
\wedge {}\psi^{\bot} =} & & \nonumber \\
& & N^{-1} (\hat{D}_{0} D - D\hat{D}_{0})\psi^{{\rm a}} -
\left(\Re^{{\rm a}}\,_{{\rm b}\bot} 
+ a^{{\rm a}} K_{{\rm b}} - a_{{\rm b}} 
K^{{\rm a}}\right) \wedge\psi^{{\rm b}}\, .
\label{Dsquared}
\end{eqnarray}
Since each side of the above equation is independent 
of the other, we get two relations. The first yields
the ``dynamical'' {\sc gcm} equation 
(\ref{triadGCM}d). The other equation we get from 
(\ref{Dsquared}) is the desired one (\ref{commutatorhatD_0andD}).

In the expression (\ref{commutatorhatD_0andD}) the curvature 
term can be eliminated in favor of extrinsic-curvature and 
acceleration ($a_{\rm b} = D_{\rm b} \log N$) terms. 
Indeed, using the {\sc gcm} equation 
(\ref{GCMe's}c), one may derive the following identity:
\begin{equation}
  \Re_{{\rm a b}\bot {\rm d}} 
+ 2 a_{[{\rm a}} K_{{\rm b}]{\rm d}} = 
  N^{-1} [D_{\rm a} (N K_{({\rm d b})}) 
+ D_{\rm d} (N K_{({\rm a b})}) 
- D_{\rm b}(N K_{({\rm d a})}) 
- D_{\rm d} (N K_{\rm b a})]
{\,} .
\label{AppendixBidentity}
\end{equation}
Remarkably, on the way to this result all of the torsion 
terms stemming from (\ref{GCMe's}c) conspire in such a way 
that only the symmetric piece of the extrinsic curvature 
tensor appears in all but the last term above. 
If desired, this identity may be 
inserted into the result (\ref{commutatorhatD_0andD}) for 
the commutator. As an application of these considerations, 
consider a triad-valued 0-form $f_{\rm b} = 
e_{\rm b}{}^{i} f_{i}$ obtained by ``soldering'' the free 
index of a $\Sigma$ 1-form $f_{i}$. With (\ref{AppendixBidentity}) 
and the evolution rules (\ref{indextriadevolution}) in hand, a 
careful examination of the expression 
$e^{\rm b}{}_{j} {\,} i_{\partial_{i}}([\hat{D}_{0}, D]f_{\rm b})$ 
(note the ``hook'' with $\partial_{i}$) yields the tensor-index
form of the commutator. Namely,
\begin{equation}
[\hat{\partial}_{0}, \bar{\nabla}_{i}]f_{j}
= [\bar{\nabla}_{i}(N K_{(jk)})
+ \bar{\nabla}_{j}(N K_{(ik)}) 
- \bar{\nabla}_{k} (N K_{(ij)})] f^{k} 
{\,} .
\end{equation}
This commutator, expanded out and written in terms of the 
acceleration 1-form $a_{i} = \bar{\nabla}_{i} \log N$, 
is featured prominently in the works of {\sc aacy}.


\section{Time derivative of the triad curvature}

Recall that in index notation we have the following definition 
of the triad curvature 2-form: 
  $R_{{\rm a}ij} \equiv - \frac{1}{2} 
   \epsilon_{{\rm a}{\rm b}{\rm c}} 
   R^{{\rm b}{\rm c}}\,_{ij}$, 
where the $R^{{\rm b}{\rm c}}\,_{ij}$ are the mixed 
triad-coordinate components of the $\Sigma$ Riemann tensor. 
We begin by using the {\sc gcm} equation 
(\ref{triadGCM}c) to derive the result
\begin{equation}
    \hat{D}_{0} R_{{\rm a}} =
  - {\textstyle \frac{1}{2}} 
    \epsilon_{{\rm a}{\rm b}{\rm c}} 
    D (N \Re^{{\rm b}{\rm c}}\,_{\bot} 
  + 2 N a^{{\rm b}} K^{{\rm c}}) 
{\,} .
\label{firstRdot}
\end{equation}
This identity is essentially the 
$({\cal D} \Re^{{\rm a}}\,_{{\rm b}})_{\bot}$ piece of the full 
(or ``un-contracted") Bianchi identity ${\cal D}\Re^{A}{}_{B} = 0$ 
for the spacetime curvature 2-form. Indeed, another method to 
derive (\ref{firstRdot}) involves a $3+1$ form decomposition of 
the full Bianchi identity. 

We now derive several auxiliary results which will be used with
(\ref{firstRdot}) to get the appropriate expression for 
$\hat{D}_{0} R_{{\rm a}}$. First, using the explicit expression
(\ref{pi_a}) for $\pi_{{\rm a}}$ along with the {\sc gcm} 
equation (\ref{triadGCM}b), we obtain
\begin{equation}
  {\textstyle \frac{1}{2}}(\Re_{\bot {\rm a}}{}^{\rm bc})
  e^{\star}_{\rm bc} = 
- 2 {}^{\star}\! D{}^{\star}\!\pi_{{\rm a}} 
- (D^{\rm b}H)e^{\star}_{\rm ab}
- (D^{\rm b}T^{\rm c}{}_{\rm a})e^{\star}_{\rm bc}
{\,} .
\label{Astep}
\end{equation}
Second, performing some permutation-symbol gymnastics and
appealing to the identity (\ref{indexT_is_antiG}) we find
\begin{equation}
  {\textstyle \frac{1}{2}}
  \epsilon_{{\rm a}{\rm b}{\rm c}}
  \Re^{{\rm b}{\rm c}}{}_{\bot} 
- {\textstyle \frac{1}{2}}
  (\Re^{{\rm b}{\rm c}}{}_{\bot{\rm a}})
  e^{\star}_{\rm bc} =  
- ({}^{\star}{}^{\parasmall}\! 
  {\sf G}^{\rm b})e^{\star}_{\rm ab}
+ \epsilon_{\rm abc} T^{\rm b}{}_{\rm d} 
  a^{\rm d} e^{\rm c}
{\,} .
\label{Bstep}
\end{equation}
Third, with the definition (\ref{definition_pi_ab}) we 
straightforwardly write 
\begin{equation}
\epsilon_{\rm abc} a^{\rm b} K^{\rm c} = 
- 2\epsilon_{\rm abc} a^{\rm b}\, {}^{\star}\!\pi^{\rm c} 
+ H a^{\rm b} e^{\star}_{\rm ab} 
- \epsilon_{\rm abc} 
  T^{\rm b}{}_{\rm d} a^{\rm c} e^{\rm d}
{\,} .
\label{Cstep}
\end{equation}
Notice the $c \leftrightarrow d$ antisymmetry present in the 
last terms of (\ref{Bstep}) and (\ref{Cstep}) [we have {\em not} 
employed the $e^{\star}_{\rm ab}$ notation in (\ref{Bstep}) in 
order to highlight this antisymmetry]. Finally, we {\em define} 
the triad-valued 1-form ${\sf T}_{\rm a}$ with the seemingly 
strange equation
\begin{equation}
- {\textstyle \frac{1}{2}}(\Re_{\bot{\rm a}}{}^{\rm bc}
- \Re^{\rm bc}{}_{\bot{\rm a}} )e^{\star}_{\rm bc} =
  {\sf T}_{\rm a} 
+ \left(D^{\rm b} T^{\rm c}{}_{\rm a}
- a^{\rm b} T^{\rm c}{}_{\rm a}
+ {\textstyle \frac{1}{2}} a_{\rm a} T^{\rm bc}
  \right)e^{\star}_{\rm bc} 
{\,} .
\label{Dstep}
\end{equation}
We have thus defined ${\sf T}_{\rm a}$, because we demand 
that the 
tensor $C_{ij}$ (\ref{C_ij}) stemming from ${\sf T}_{\rm a}$ 
satisfies the property (\ref{Laplacian_on_T}). This property 
would seem essential for the proof of equivalence given in 
$\S$ V to go through. Although it may yet be possible to 
modify the definition of ${\sf T}_{\rm a}$ in an advantageous 
way, the definition that we adopt here provides us with a 
consistent formalism. Using the {\sc gcm} splitting equations 
(\ref{GCMe's}b,c), we find\footnote{Alternatively, we 
can write
$$
  {\sf T}_{\rm a} = {\textstyle \frac{1}{4}}\left[
  N D_{\rm a} (N^{-1} T^{\rm bc}) 
- 2 N D^{\rm b} (N^{-1} T^{\rm c}{}_{\rm a})\right]
  e^{\star}_{\rm bc}
{\,} .
$$
This results allows one to easily obtain the explicit 
expression for $C_{ij}$ given in Eq.~(\ref{C_ij}).}
\begin{equation}
{\sf T}_{\rm a} = 
  {\textstyle \frac{1}{2}}\delta_{\rm ab}
  N [{}^{\star}\!
  d(N^{-1} T)]e^{\rm b}
- \delta_{\rm ab} N\, {}^{\star}\!D {}^{\star}
  (N^{-1} e^{\rm b} \wedge {}^{\star}T) 
 {\,} .
\label{torsionterms}
\end{equation}
as the abstract expression for this 1-form.

Now add together the equations (\ref{Astep}), (\ref{Bstep}), 
(\ref{Cstep}), and (\ref{Dstep}) and then perform a short
calculation (taking advantage of the aforementioned 
$c \leftrightarrow d$ antisymmetry) in order to reach
the following result:
\begin{eqnarray}
{\textstyle \frac{1}{2}}\epsilon_{{\rm a}{\rm b}{\rm c}}
  (\Re^{{\rm b}{\rm c}}{}_{\bot} + 2 a^{{\rm b}} K^{{\rm c}}) 
& = &
  {\sf T}_{{\rm a}}
- ({}^{\star}{}^{\parasmall}\!{\sf G}^{\rm b})
  e^{\star}_{\rm ab} 
- 2{}^{\star}\! D {}^{\star}\!\pi_{{\rm a}}
\nonumber \\
& &
- 2 \epsilon_{{\rm a}{\rm b}{\rm c}} a^{{\rm b}}{\,} 
  {}^{\star}\!\pi^{{\rm c}}
+ (a^{\rm b} H - D^{\rm b} H) e^{\star}_{\rm ab}
+ a_{\rm a} {}^{\star} T
{\,} .
\label{Estep}
\end{eqnarray}
Finally, plugging (\ref{Estep}) into (\ref{firstRdot}) we arrive
at
\begin{eqnarray}
\hat{D}_{0} R_{{\rm a}} & = &
- N {}^{\star}\! D
  {}^{\star}{}^{\parasmall}\!{\sf G}_{{\rm a}}
- e^{\star}_{\rm ab} \wedge D
  {}^{\star}(N {}^{\parasmall}\!{\sf G}^{\rm b})
- D(N{\sf T}_{{\rm a}}) 
- 2 N \Delta \pi_{{\rm a}}
+ 2N a \wedge {}^{\star}\! D {}^{\star}\!\pi_{{\rm a}}
\nonumber \\
& &
+ 2 \epsilon_{{\rm a}{\rm b}{\rm c}} 
  D(N a^{{\rm b}}{\,} 
  {}^{\star}\!\pi^{{\rm c}})
+ e^{\star}_{\rm ab} \wedge
  D(N H a^{{\rm b}})
- e^{\star}_{\rm ab} \wedge
  D(N D^{\rm b} H)
- D(N a_{\rm a} {}^{\star} T)
{\,} .
\label{mainRdot}
\end{eqnarray}
To get this identity, we have introduced the triad-covariant
Laplacian $\Delta = 
({}^{\star}\!D {}^{\star}\! D - D{}^{\star}\!D{}^{\star})$
on triad-valued 2-forms
at the price of an appeal to (\ref{3+1etf}a). Also, in this
expression we consider $T$ as a short-hand for  $2 e^{\rm a} 
\wedge {}^{\star}\!\pi_{\rm a}$ 
[cf.~Eq.~(\ref{T_from_pi})].
We remark that other useful expressions for the final form of 
$\hat{D}_{0}R_{{\rm a}}$ are certainly possible. However,
although we would, of course, prefer to incorporate the 
last torsional term appearing in (\ref{mainRdot}) into 
the definition of ${\sf T}_{\rm a}$ (in order to obtain a more
compact expression for the final result), this would seem not 
to be permissible. Indeed, were we to alter the definition of 
${\sf T}_{\rm a}$ in this manner, we would lose the crucial 
condition (\ref{Laplacian_on_T}). More 
precisely, we would have to replace the {\rm (l.h.E.T.)}$_{0}$ 
in that equation with {\rm (l.h.E.T.)}$_{1}$. This in turn 
would seem to spoil the proof of equivalence 
between our {\sc df} system of equations and the original Einstein 
equations.


\end{document}